\documentclass[]{aa}  
\usepackage{graphicx}
\usepackage[varg]{txfonts}
\usepackage{hyperref}
\usepackage{xcolor}
\usepackage{float}
\usepackage{natbib}
\usepackage{lscape}
\bibpunct{(}{)}{;}{a}{}{,}
\usepackage{epstopdf}
\usepackage{subcaption}
\hypersetup{draft}

\begin{document} 

\title{Abundances of disk and Bulge giants from high-resolution optical spectra\thanks{Based on observations made with the Nordic Optical Telescope (programs 51-018 and 53-002), operated by the Nordic Optical Telescope Scientific Association at the Observatorio del Roque de los Muchachos, La Palma, Spain, of the Instituto de Astrofisica de Canarias, and on spectral data retrieved from PolarBase at Observatoire Midi Pyrénées. Tables \ref{tab:basicdata_sn}-\ref{tab:abundances_bulge} are only available in electronic form at the CDS via anonymous ftp to cdsarc.u-strasbg.fr (130.79.128.5) or via \href{http://cdsweb.u-strasbg.fr/cgi-bin/qcat?J/A+A/}{http://cdsweb.u-strasbg.fr/cgi-bin/qcat?J/A+A/}}}
\subtitle{III. Sc, V, Cr, Mn, Co, Ni}
\author{M.~Lomaeva\inst{1} 
\and H.~J\"onsson\inst{1} 
\and N.~Ryde\inst{1} 
\and M.~Schultheis\inst{2} 
\and B.~Thorsbro\inst{1} 
}
\institute{Lund Observatory, Department of Astronomy and Theoretical Physics, Lund University, Box 43, SE-221 00 Lund, Sweden\\\email{lomaevamaria@gmail.com} \and
Observatoire de la Cote d'Azur, Boulevard de l'Observatoire, B.P. 4229, F 06304 NICE Cedex 4, France
}

\date{Received 2018; accepted 2019}

% \abstract{}{}{}{}{} 
% 5 {} token are mandatory
 
  \abstract
   {%The formation and evolution of the Galactic Bulge and the Milky Way is still a debated subject. Observations of, e.g., the X-shaped Bulge, cylindrical stellar motions, and presumed existence of a fraction of young stars in the Bulge have suggested that it formed through secular evolution of the disk and not through gas dissipation and/or mergers, as thought previously.
   Recent observations of the Bulge, e.g., its X-shape, cylindrical stellar motions, and a potential fraction of young stars propose that it formed through secular evolution of the disk and not through gas dissipation and/or mergers, as thought previously.
   }
   {%The goal is to measure abundances of six iron-peak elements (Sc, V, Cr, Mn, Co and Ni) in the local thin and thick disks as well as the Bulge. These can provide additional observational constraints for Galaxy formation and chemical evolution models and help to understand whether the Bulge has emerged from the (thick) disk or not.
  We measure abundances of six iron-peak elements (Sc, V, Cr, Mn, Co and Ni) in the local thin and thick disks as well as the Bulge to provide additional observational constraints for Galaxy formation and chemical evolution models.
  }
   {%We use high-resolution optical spectra of 291 K giants in the local disk mostly obtained by the FIES at NOT (signal-to-noise (S/N) ratio of 80-100) and 45 K giants in the Bulge obtained by the UVES/FLAMES at VLT (S/N ratio of 10-80). The abundances are measured using SME. Additionally, we apply NLTE corrections to the [Mn/Fe] and [Co/Fe] ratios. The thin and thick were separated according to their metallicity, [Ti/Fe]-ratios as well as proper motions and the radial velocities from Gaia DR2.
   We use high-resolution optical spectra of 291 K giants in the local disk mostly obtained by the FIES at NOT (signal-to-noise (S/N) ratio of 80-100) and 45 K giants in the Bulge obtained by the UVES/FLAMES at VLT (S/N ratio of 10-80). We measure abundances in SME and apply NLTE corrections to the [Mn/Fe] and [Co/Fe] ratios. To discriminate between the thin and thick, we use stellar metallicity, [Ti/Fe]-ratios, and kinematics from Gaia DR2 (proper motions and the radial velocities).
   }
   {%The trend of [V/Fe] vs. [Fe/H] shows a separation between the disk components, being more enhanced in the thick disk. Similarly,  the [Co/Fe] vs. [Fe/H] trend shows a hint of an enhancement in the local thick disk.  The trends of V and Co in the Bulge  appear to be even more enhanced, although within the uncertainties. The decreasing [Sc/Fe] ratio with increasing metallicity is observed in all the components, while our [Mn/Fe] ratio steadily increases with increasing metallicity in the local disk and the Bulge instead. For Cr and Ni we find a flat trend following iron for the whole metallicity range in the disk and the Bulge. The [Ni/Fe] ratio appears slightly overabundant in the thick disk and the Bulge compared to the thin disk, although the difference is minor.
   The observed disk trend of V is more enhanced in the thick disk, while the Co disk trend shows a minor enhancement in the thick disk.  The Bulge trends of V and Co appear even more enhanced w.r.t. the thick disk, but within the uncertainties. The [Ni/Fe] ratio seems slightly overabundant in the thick disk and the Bulge w.r.t. the thin disk, although the difference is minor. The disk and Bulge trends of Sc, Cr and Mn overlap strongly.
   } 
   {The somewhat enhanced  [(V,Co)/Fe] ratios observed in the Bulge suggest that the {\it local} thick disk and the Bulge might have experienced different chemical enrichment and evolutionary paths. However, we are unable to predict the exact evolutionary path of the Bulge solely based on these observations. Galactic chemical evolution models could, on the other hand, provide that using these results. }


   \maketitle
%
%________________________________________________________________
\section{Introduction}

One of the most important questions in contemporary astrophysics concerns the formation and evolution of galaxies \citep{bh16}. The Milky Way is a typical barred spiral galaxy and is the only such galaxy where we can resolve individual stars for a detailed analysis. To fully understand the formation of the Milky Way, it is essential to understand the formation of its central region, the Bulge. The view on the origin of the Bulge has changed drastically over the past years. Previously, the Bulge was thought to belong to the group of spherically shaped classical bulges formed through dissipation of gas \citep[e.g.,][]{eggen62, chiappini97, micali13} or merging events according to the $\Lambda$CDM theory \citep[e.g.,][]{abadi03, scan03}. However, this view has been challenged by new observations revealing properties of the Milky Way Bulge that are not typical for classical bulges.  

These new observations include  kinematic observations from the Apache Point Observatory Galactic Evolution Experiment (APOGEE) \citep{zasowski:16,ness16}, 
the X-shape of the Bulge which represents the inner, 3-D part of the Galactic bar \citep{mcw_z10, nataf10, wegg13}, cylindrical rotation of the Bulge stars from the BRAVA \citep{kunder12} and ARGOS \citep{ness13b} surveys and a population of presumably young stars found in the Bulge. \cite{loon03} found stars as young as $\lesssim$ 200 Myr across the inner Bulge. More recent studies of blue stragglers by \cite{clarkson11} and microlensed dwarfs by \cite{bens17} have suggested that there is an,  at least, fractional stellar component in the Bulge with ages $\lesssim$ 5 Gyr. A generation with ages $\lesssim$ 5 Gyr was also discovered by \cite{bernard18} who studied deep colour-magnitude diagrams of four Bulge fields.% obtained with the Hubble Space Telescope.

The new discoveries point towards the Bulge being dynamically formed and having the boxy/peanut (b/p) shape which also has been observed in many other spiral galaxies (\citealt{lutticke00, kormendy04}). Several N-body simulations of the evolution of spiral galaxies have managed to reproduce the b/p Bulge shape through secular disk instabilities  \citep[e.g.,][]{c_s81, a05, mv06}. In such simulations, stellar bars experience one or even multiple buckling instabilities which result in a boxy or peanut-shaped structure depending on the viewing angle (e.g. \citealt{dimatteo14, fragkoudi18}). Moreover, bulges formed through disk instabilities,  being of disk origin \citep{fragkoudi18}, can and, in fact, might be expected to contain young stars  \citep{gonz16}.

However, the structure of the Bulge appears to be non-homogeneous. Different kinematic properties measured by APOGEE of populations of different metallicities \citep{zasowski:16} provide evidence for different evolutionary histories of these populations, and the metallicity distribution maps found by the GIBS survey \citep{zoccali:17} and APOGEE \citep{garcia:18} show the presence of different populations. \cite{bab10} studied kinematics (radial velocities and proper motions) and metallicities of a sample of Bulge stars spread out over different latitudes and concluded that metal-rich stars can be associated with a barred population, while the metal-poor can be associated with a spheroidal component or potentially the inner thick disc. Additionally, \cite{ness13b} and \cite{vasc13}, who also studied chemo-dynamical properties of Bulge stars, arrived at the conclusion that  metal-rich stars in their sample belong to the X-shaped Bulge, whereas the metal-poor do not. These results and some dynamical simulations \citep[e.g.,][]{dimatteo14} allow the existence of a minor spherical component, while most of the mass in the Bulge originates from the disk. \cite{shen10} modelled the cylindrical rotation in the Bulge and concluded that the mass of the spheroidal component can at largest be 8\% of the disk mass in order to reproduce the BRAVA observations.

Evidently, the formation of the Galaxy and the Bulge is a complex matter and is still not fully understood as no successful fully self-consistent models for the Bulge in the cosmological framework are available today \citep{barb18}. 

In order to observationally constrain potential formation routes, abundance ratios can be measured to, e.g.,  constrain the Star Formation Rate (SFR) and Initial Mass Function (IMF) of a population. To quantify the SFR and IMF  in the Bulge, [$\alpha$/Fe] vs. [Fe/H] trends have been used extensively \citep{matteucci_book}. E.g., \citet{bensby:10} studied $\alpha$-trends of K giant stars in the Bulge and inner thick disk (galactocentric radius of 4-7 kpc) and find chemical similarities, also between the inner and local thick disks. Later,  \cite{bens17} compare their $\alpha$-trends of dwarf stars in the local thick disk and Bulge and find that there is no significant variation in IMF, however, the [$\alpha$/Fe] ``knee'' appears at slightly higher metallicities in the Bulge than in the local thick disk. They note that this question should be further investigated by analysing larger stellar samples. 

In the first two papers in this series, \cite{jonsson17_d, jonsson17_b} worked on the same spectra as are used in this paper, of disk and Bulge giants and determined their $\alpha$-abundances, which led to the  conclusion that the Bulge generally follows the thick disk trend, again indicating similar chemical evolution histories. These results could probably indicate a slightly higher SFR in the Bulge, too, since the Bulge trends of Mg, Ca and Ti trace the upper envelope of the thick disk trends. 

\cite{johnson14} examined $\alpha$ and iron-peak abundances in Bulge giants and arrived at a different result: while no special IMF is required to reproduce the Bulge trends, they conclude that the Bulge and the thick disk have experienced different chemical enrichment paths. This conclusion was drawn from the enhanced abundance trends, in particular of the iron-peak elements such as Co, Ni and Cu, in the Bulge compared to the thick disk.  In the analysis, however, their reference sample consisted of disk dwarfs, and this sort of comparison using dwarfs and giants is likely to be affected by potential systematic offsets \citep[see][]{melendez08, alves-brito10}. 

Yet, \citet{bens17} did not find such an enhancement for Ni in the Bulge. The trend of a different iron-peak element, Mn, has also been been a subject to disagreement in terms of the separation between the thin and thick disk components \citep{feltz07,bb}.

From the APOGEE survey there have been two papers discussing abundance trends in the Bulge: \citet{schultheis17} and \citet{zasowski2018}. Regarding iron-peak elements, they both investigate Cr, Mn, Co, and Ni, and \citet{zasowski2018} compare to similarly analysed stars in the local disk. For Cr, and possibly Co, they find differences in the local and Bulge trends, while the trends for Mn, and Ni seem very similar in the two populations.

Clearly, the literature studies have not arrived at a common conclusion as some are finding similarities and others find differences between the evolutionary paths of the disk and the Bulge. A manual, homogeneous spectroscopic analysis of the same type of stars in these Galactic components, examining other elements than the extensively studied $\alpha$-elements, can give new insights into the question of whether or not the Bulge emerged from the disk. 

Iron-peak elements, as mentioned above, are suitable for such Galactic archaeology studies, being able to probe the chemical enrichment history. In this paper, we examine iron-peak elements with atomic numbers 21 $\leq Z \leq $  28, i.e., Sc, V, Cr, Mn, Fe, Co and Ni\footnote{Titanium is often defined as an $\alpha$-element but sometimes also as an iron-peak element \citep[e.g.,][]{sneden16}}.   Great progress has been made in understanding their nucleosynthetic sites and production, but it is still a subject for debate.  Especially for Sc, V, and Co, the observational trends can not be reproduced with galactic chemical evolution models, and for Mn, the uncertainties in the observed trends need further investigations. For Cr and Ni the situation is, however, better. 

Iron-peak elements are synthesised both in thermonuclear and core-collapse SNe. Similarly as for $\alpha$-elements, SNe II are presumed to be the main source of Sc (\citealt{clayton}, hereafter C03). Whereas V is thought to be mostly created in SNe Ia (C03), Cr is created in SNe Ia and SNe II in comparable amounts (C03). This is also true for  the heavier iron-peak elements, Fe, Co and Ni, eventhough they are formed through other nucleosynthetic channels (C03). The understanding of the synthesis of Mn is more complicated with possible sites in both SNe Ia and SNe II. For SNe Ia, the amount of Mn  yields depend on the properties of the progenitor white dwarf \citep[e.g.,][]{nomoto97, yama15}.

In this work, we continue our study in \cite{jonsson17_d, jonsson17_b}, where abundances of 
O, Mg, Ca and Ti were determined, and examine the iron-peak elements Sc, V, Cr, Mn, Co and Ni conducting a homogeneous giant-giant comparison of the local disks and the Bulge.

%__________________________________________________________________
\section{Observations}
\subsection{Solar neighbourhood sample}

In the solar neighbourhood sample there are 291 K giants. %The exact distance to these stars is not known, but they are located inside a sphere with a radius of 1-2 kpc to which the solar neighbourhood is confined \citep[e.g.,][]{klement10}. 
Most of these stars were observed using the spectrometer FIES \citep{telting14} installed on the Nordic Optical Telescope (NOT) in May-June 2015 under programme 51-018 (150 stars) and June 2016 under programme 53-002 (63 stars). 41 spectra were taken from \cite{thygesen12} in turn from FIES/NOT, 18 were downloaded from the FIES archive, and 19 spectra were taken from the NARVAL\footnote{ Mounted on Telescope Bernard Lyot (TBL)} and ESPaDOnS\footnote{ Mounted on the Canada--France--Hawaii Telescope (CFHT)} spectral archive in the PolarBase data base \citep{petit2014}. 

The  resolving power of the FIES spectra is $R\sim$67000 and for PolarBase it is $R\sim$65000. The entire optical spectrum was covered by these instruments, but the wavelength region was restricted to 5800-6800 \r{A} in order to match the wavelength region of the Bulge spectra, providing a more homogeneous analysis.

The majority of the observed stars are quite bright, see Table \ref{tab:basicdata_sn} (Online material), and the corresponding observing time is, in many cases, of the order of minutes. For the 213 FIES stars, the `expcount'-feature was used during the observations which allows to abort the exposure when a certain CCD count level has been obtained. The signal-to-noise (S/N) ratios of the FIES spectra are generally high: between 80 and 120  per data point in the reduced spectrum, see Table \ref{tab:basicdata_sn} (Online material); roughly the same applies to the spectra from PolarBase. Spectra from \cite{thygesen12} have a lower S/N ratio of about 30-50. The S/N ratios were measured by the IDL-routine \texttt{der\_snr.pro}\footnote{See \texttt{stecf.org/software/ASTROsoft/DER\_SNR}} and are listed in Table \ref{tab:basicdata_sn} (Online material).

The reduction of the FIES spectra was preformed using the standard FIES pipeline. The spectra from \cite{thygesen12} and PolarBase were already reduced and ready to use.  An initial, rough normalisation of all of the spectra was done with the IRAF task \texttt{continuum}. However, in the analysis, the continuum is normalised more carefully by fitting of a straight line to continuum regions in every (short) wavelength window examined. No removal of atmospheric telluric lines or subtraction of the sky emission lines has been attempted, those were avoided instead by comparison to the (radial velocity shifted) telluric spectrum of the Arcturus atlas of \citet{hink00}.

Most of the stars from the thick disk lie at a distance of $\sim$45-2000 pc from the Sun, whereas the majority of the thin disk stars are located  $\sim$30-1000 pc away using the estimations from \cite{mcmillan18}. The separation of the disk components is discussed in Section \ref{disk_sep}.

\begin{figure*}[!th]\centering  
\includegraphics[width=\linewidth]{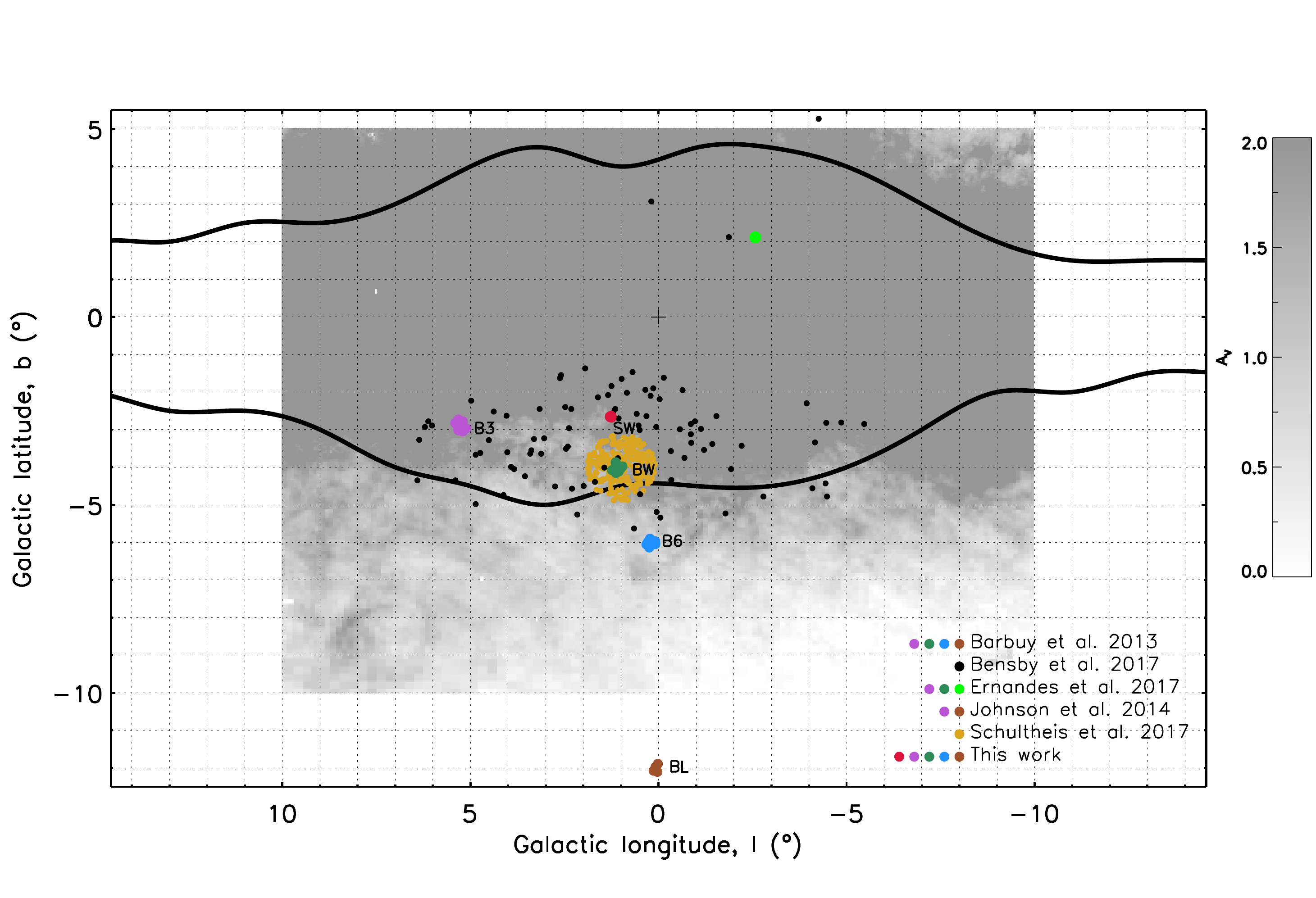} 
\caption[Bulge fields analysed]{The map of the Galactic Bulge showing the five analysed fields (SW, B3, BW, B6, and BL). The positions of the microlensed dwarfs from \cite{bens17} and fields analysed in \cite{barb13}, \cite{ernandes18}, \cite{johnson14} and \cite{schultheis17} are also marked in the figure. The dust extinction towards the Bulge is taken from \cite{gonz11, gonz12} scaled to optical extinction \citep{cardelli89}. The scale saturates at $A_\mathrm V = 2$. The COBE/DIRBE contours of the Galactic Bulge are taken from \cite{weiland1994}}.
\label{bulge_field}\end{figure*}

\subsection{Galactic Bulge sample}
The Galactic Bulge sample consists of 45 K giants, see Table \ref{tab:basicdata_bulge}. The spectra were obtained using the spectrometer FLAMES/UVES installed at the VLT. In total, with the original aim of attempting to investigate gradients,  five fields  were observed: SW, BW, BL, B3 and B6. The naming convention of these fields is based on that of \citet{lec07}; B3 means a field at $b=-3^\circ$, B6 a field at $b=-6^\circ$ along the Galactic minor axis, SW and BW refers to the Sagittarius Window and Baade's Window, respectively, and BL to the Blanco field. The definition of the extent of the Bulge is a bit blurred \citep[see discussion in][]{barb18}, but with the \citet{barb18} definition of the Bulge being the region on the sky within $b=\pm10^\circ$ and $l=\pm10^\circ$, the Blanco field is on the verge of being a `bulge' field, but we include it, nevertheless, in our Bulge compilation, to be consistent with e.g. \citet{lec07} and \citet{johnson14}. The fields are marked in Figure \ref{bulge_field}, together with the outline of the Galactic Bulge \citep{weiland1994}, the positions of the microlensed dwarfs from \cite{bens17} and fields analysed in \cite{barb13}, \cite{ernandes18}, \cite{johnson14} and \cite{schultheis17}, studies that we will discuss later in Section \ref{b_comp}. Due to the dust extinction it is difficult to observe Bulge stars in the optical wavelength range, and for this reason it was only possible to observe stars in the outer part of the Bulge. Observations of stars in the immediate centre require observations in the IR. However, determination of stellar parameters for such observations is still difficult (e.g. \citealt{rich12}) but progress is being made \citep[e.g.,][]{ryde16, schultheis16, rich17, nandakumar18}. %In the five regions analysed here, the extinction is less severe compared with the surroundings.  

The 34 stars in the fields B3, BW, B6 and BL were observed in May-August 2003-2004, while the 11 stars in the SW field were observed in August 2011 (ESO program 085.B-0552(A)). With FLAMES/UVES it is possible to observe seven stars in each pointing. Depending on the extinction, each setting required an integration time of 5-12 hours. The achieved S/N ratios of the spectra in this sample are significantly lower than  those for the stars in the solar neighbourhood sample, ranging between 10 and 80. The resolving power of the Bulge spectra is $R \approx$ 47000 and  the wavelength coverage is between 5800 and 6800 \r{A}.

All of the stars in the Bulge sample apart from one have a parallax uncertainty above 20\% as measured by Gaia DR2 \citep{gaia16,gaia18} resulting in rather uncertain distance  estimations. Nevertheless, \cite{bailerjones18} estimate that most of these stars lie 4-12 kpc away, with a tendency to lower values. This means that our stars are spread out over the area occupied by the Bulge/bar \citep{wegg15}.  We expect higher reliability of distances for our Bulge stars from coming Gaia data releases.
%will investigate further the reliability of the Bulge distances in the future.

%\hspace{5cm}

%__________________________________________________________________
\section{Analysis}
The analysis in this investigation is performed in the same manner as in first papers of this series \citep{jonsson17_d, jonsson17_b}. In general, every observed spectral line of interest is fitted, by means of a $\chi^2$ minimization, with synthetic spectra modelling the line strengths and profiles, for a given set of stellar parameters, as determined in \citet{jonsson17_d, jonsson17_b}. The atomic line data needed is described in Section \ref{sec:linedata} and the spectral synthesizing tool, Spectroscopy Made Easy (SME), in Section \ref{sec:sme}.  The  model atmospheres that are used and interpolated for in SME is described in Section \ref{sec:model_atm}. We also discuss the previously determined photospheric stellar parameters for our stars in Section \ref{sec:stel_par}. The method used for discriminating between the thin and thick disk is described in Section \ref{disk_sep}.

\subsection{Line data}\label{sec:linedata} 
The line data used in the determination of chemical abundances for all elements apart from Sc were taken from the Gaia-ESO line list version 5 (\citealt{heiter15}, Heiter et al., in prep.). For odd-Z elements\footnote{Here: Sc, V, Co and Mn}, one has to consider the hyperfine splitting (hfs) of atomic energy levels, that will have a de-saturating effect on  strong atomic lines \citep{pro00,thorsbro18}. Since the hfs components for Sc were not present in the Gaia-ESO line list, they were instead taken from the updated version of the VALD line list \citep{vald99, vald17}, see Table \ref{tab:linedata}.  All lines used, apart from those from Sc, are from the neutral species. The Sc lines are from \ion{Sc}{ii}.

We have taken several precautions to try to avoid blended lines: first we followed the recommendations from Gaia-ESO \citep{heiter15}, then we carefully scanned our selected lines for visible blends in all stars, and finally we determined abundances for all lines individually to make sure they reproduce the same abundance trends with metallicity.

\subsection{Spectral line synthesis}\label{sec:sme}
For abundance measurements, we used the spectral line synthesiser Spectroscopy Made Easy (SME; \citealt{sme96, sme17}).  SME can simultaneously fit global stellar photospheric parameters and/or -- as is done in this work -- abundances based on user defined line masks covering spectral lines of interest.  To do so, synthetic model spectra of different abundances are synthesised on the fly, and the best fit of the line profile to the observed data is found utilising the $\chi^2$ minimisation method as described in \citet{marquardt1963}. For the elements with several lines of suitable strength available (see Table \ref{tab:linedata}), all lines were fitted simultaneously, but investigatory individual fits were also done  to make sure that no line was systematically deviating from the others. The other abundances in the calculation of the synthetic spectra are solar, by default from \cite{grev07}, scaled with metallicity unless defined otherwise, which we do for the $\alpha$ elements, determined in \citet{jonsson17_b}. 

\subsection{Model atmospheres}\label{sec:model_atm}
 We used the grid of  LTE\footnote{Local Thermodynamic Equilibrium} MARCS models \citep{gustaf08} supplied with SME, which consists of a subsample of the models available on the MARCS webpage\footnote{See \texttt{marcs.astro.uu.se}}. The model grid is $\alpha$-enhanced according to the standard MARCS-scheme with [$\alpha$/Fe] = +0.4 for [Fe/H] < $-$1, [$\alpha$/Fe] = 0.0 for [Fe/H] > 0, and linearly falling in-between, while the other abundances are solar values simply scaled with metallicity. The models are spherically symmetric for $\log g<$ 3 and plane parallel otherwise.
For the spectral synthesis, SME interpolates in this grid of model atmospheres, keeping the non-fitted abundances consistent with the grid of model atmospheres. While this on-the-fly spectral synthesis is made under the assumption of LTE, we later add available NLTE abundance corrections for the Co and Mn trends. 

\subsection{Photospheric stellar parameters}\label{sec:stel_par}
All photospheric stellar parameters used here were determined in \cite{jonsson17_d, jonsson17_b}. Briefly, they were estimated by simultaneously fitting a synthetic spectrum  using the same version of SME and model atmosphere grid for unsaturated and unblended \ion{Fe}{i}, \ion{Fe}{ii} and \ion{Ca}{i} lines as well as $\log g$ sensitive  \ion{Ca}{i} wings, while $T_{\textnormal{eff}}$, $\log g$, [Fe/H], $v_{\textnormal{mic}}$ and [Ca/Fe] were set as free parameters. NLTE corrections for  \ion{Fe}{i}  from \cite{lind12} were accounted for when calculating the atmospheric parameters, but these were very small. For more details see \cite{jonsson17_d, jonsson17_b}.

Representative uncertainties in the stellar parameters for a disk star of S/N ratio $\sim$100 (per data point in the reduced spectrum as measured by \texttt{der\_snr.pro}) are $\pm$50 K for $T_{\textnormal{eff}}$, $\pm$0.15 dex for $\log g$, $\pm$0.05 dex for [Fe/H] and $\pm$0.1 km/s for $v_{\textnormal{mic}}$, but their magnitudes of course depend on the S/N ratio \citep[see Figure 2 in][]{jonsson17_d}.

\subsection{Thin and thick disk disk separation}\label{disk_sep}
The thin and thick disk stellar populations show some substantial differences in [$\alpha$/Fe] ratios, kinematics and ages.

Ages of giant stars are rather hard to determine due to the strong overlap of isochrones on the red giant branch (RGB). 
The Galactic space velocities ($U, V, W$) can be used as well, however, \cite{bens14_d} found,  that stellar ages of dwarfs act as a better discriminator between the thick and thin disk than the kinematics. Although, they note that stellar ages are often subjected to larger uncertainties. They also saw an age-[$\alpha$/Fe] relation by studying the [Ti/Fe] ratio as it shows a clear enhancement of the thick disk. %compared to [Fe/H]. 
They concluded that dwarf and sub-giant stars older than 8 Gyr exhibit higher [Ti/Fe] ratios, and Ti abundances can, therefore, be used to distinguish between the old and young stellar populations, at least in the solar neighbourhood.

For this reason and also given the clear separation between the disk components in the [Ti/Fe] vs. [Fe/H] trend  for giant stars too \citep{jonsson17_d}, we use the [Fe/H] and [Ti/Fe] abundances measured in that paper as well as the proper motions and radial velocities from Table \ref{tab:basicdata_bulge} and Gaia DR2 \citep{gaia16, gaia18} to calculate the total space velocity, $V_{\textnormal{tot}}$\footnote{$V_{\textnormal{tot}}^2 \equiv U^2 + V^2 + W^2$},  to assign the stars to either the thin or thick disks. For thick disk stars $V_{\textnormal{tot}}$ is generally higher than for the thin disk \citep{nissen04}. The kinematic data were available for 268 stars in our sample, and to convert those into (U, V, W) velocity components we used tools from the \texttt{astropy} package, and Gaia DR2 distance estimates from \cite{mcmillan18}. The remaining stars without kinematics were not considered in the separation, however, they are used in the following in the cases where the disk sample is considered as a whole.

The separation itself was done using a clustering method called Gaussian Mixture Model (GMM), which was obtained from the \texttt{scikit-learn} module written in Python \citep{scilearn} that contains the \texttt{GaussianMixture} package.  The GMM is a
parametric probability density function which is represented as a weighted sum of Gaussian component densities, i.e., the overall distribution of the data points is assumed to consists of (multi-dimensional) Gaussian sub-distributions. The parameters of the complete GMM (mean vectors, covariance matrices and mixture weights from all component densities) can be estimated from training data utilising the iterative Expectation-Maximisation (EM) algorithm.
%statistical tool which assumes that the overall distribution of the data points consists of (multi-dimensional) Gaussian sub-distributions. 
Broadly speaking, the EM algorithm first calculates the probability of the data points to belong to one of the clusters and then updates the parameters of the Gaussian sub-distributions using the estimated membership probabilities. Each iteration of the EM algorithm increases the log-likelihood of the model improving the fit to the data until it converges. However, the number of clusters has to be known in advance. Here, the number of clusters was set to two. For a thorough mathematical description see, e.g., \cite{gmm}.

Note that in the separation no S/N ratio cut was performed and abundance uncertainties were not considered.

\begin{figure*}[!th]
\centering
\includegraphics[width=0.9\linewidth]{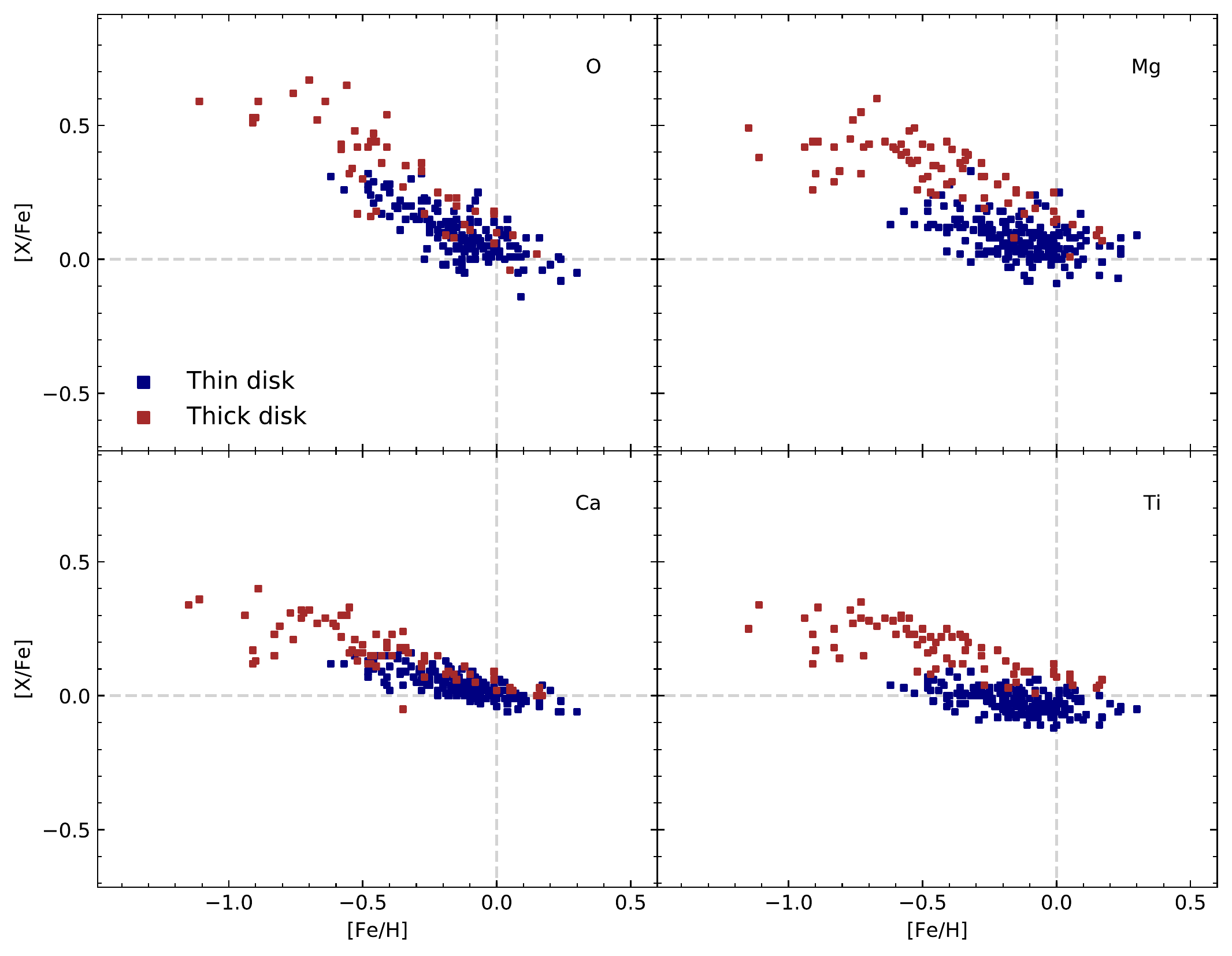}
\caption[Thin and thick disk separation]{The thin and thick disk separation (blue and red squares respectively) applied on the [(O,Mg,Ca,Ti)/Fe] vs. [Fe/H] trends from \cite{jonsson17_d}. The plots are shown in the same manner as in \cite{jonsson17_d}, i.e., with the solar abundances from \cite{asplund09}.}
\label{ti_sep}
\end{figure*}

To check the validity of the clustering, we applied the separation of the disk stars onto the trends of the $\alpha$-elements in \cite{jonsson17_d}, as shown in Figure \ref{ti_sep}. This separation was then applied to the iron-peak abundance trends.

%__________________________________________________________________

\section{Results}
The determined abundances and the obtained abundance trends of the iron-peak elements studied in the disk and the Bulge are listed in Tables \ref{tab:abundances_sn}-\ref{tab:abundances_bulge} and shown in Figures \ref{disk_plt}-\ref{bulge_plt}.
To highlight the features of the trends, we also plot the running mean and running 1$\sigma$ scatter which are shown. The number of data points in the running window was set to 29 for the thin and thick disk samples and 14 for the Bulge. Therefore, the running mean and 1$\sigma$ scatter does not cover the whole trend range. Running median was tested as well, but the results showed to be very similar so all conclusions and discussion would be qualitatively the same.

\subsection{Solar Neighbourhood}

The solar neighbourhood stellar sample consists of 291 K giants, where 268 had available kinematic data, of which 71 likely belong to the thick disk population and 197 to the thin disk. A rather clear separation between the disk components is seen in the trend of V; for Co, there is a minor overlap; and a larger overlap is seen for Sc and Ni. For the Cr and Mn disk trends, the overlap is very prominent.

\begin{figure*}[!th]
\centering  
\includegraphics[width=\linewidth]{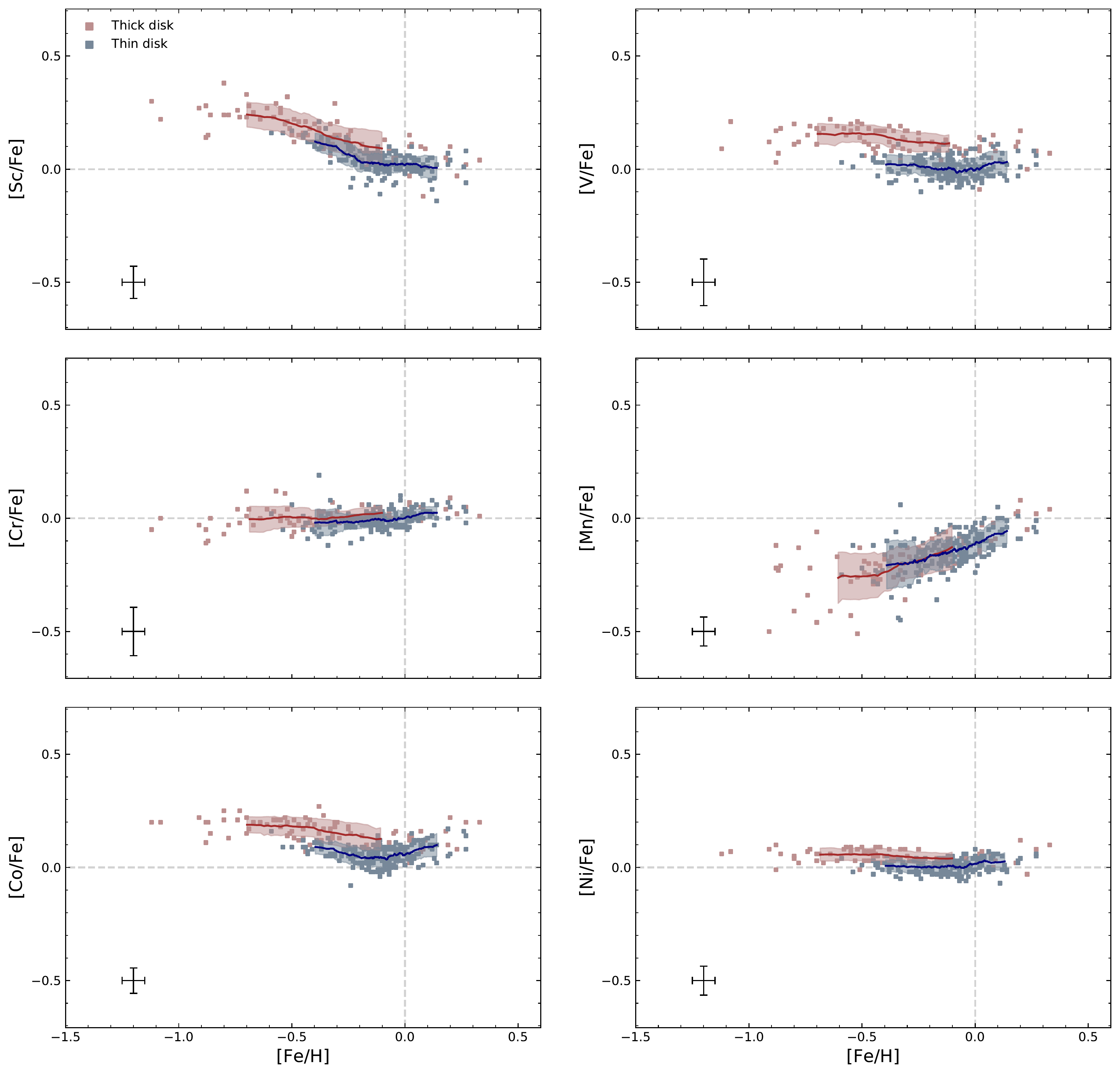} 
\caption[]{[X/Fe]  vs. [Fe/H] trends (LTE) of the examined solar neighbourhood giants: pink squares denote thick disk stars and blue squares denote thin disk stars. The running mean for the thick disk is the red solid line, for the thin disk the line is blue. The shaded areas are the running 1$\sigma$ scatter of the thick (red) and thin (blue) disks. The typical uncertainties for each element from Table \ref{mc_lbl} are also shown in the plots. In the plots we use A(Sc)$_\odot$ = 3.04 \citep{peh17_sc2}, A(V)$_\odot$ = 3.89, A(Cr)$_\odot$ = 5.62, A(Mn)$_\odot$ = 5.42,  A(Fe)$_\odot$ = 7.47, A(Co)$_\odot$ = 4.93, A(Ni)$_\odot$ =  6.20 \citep{scott15}.}
%\label{hr}
\label{disk_plt}\end{figure*}

The thick disk trend of Sc in Figure \ref{disk_plt} is somewhat enhanced at the low metallicity end reaching [Sc/Fe] $\sim$+0.25 dex (here and henceforth, we refer to the running mean when describing [X/Fe] ratios). As [Fe/H] increases, the [Sc/Fe] ratio goes down to $\sim$+0.1 dex. The average $\langle$[Sc/Fe]$\rangle$ ratio of the thick disk is +0.17 dex with a mean scatter, $\langle\sigma\rangle$, of 0.05 dex. The thin disk trend has the highest [Sc/Fe] ratio of $\sim$+0.15 dex at [Fe/H] $\sim$$-$0.4 which decreases to $\sim$0 and flattens out at [Fe/H] $\sim$$-$0.2. This plateau remains even at supersolar metallicities. The average elevation of the $\langle$[Sc/Fe]$\rangle$ ratio in the thin disk is +0.03 dex with $\langle\sigma\rangle$ = 0.04 dex. In the region where the thin and thick disk overlaps, the thick disk trend is enhanced compared to the thin disk, but the uncertainties of the trends highly overlap.
 
The [V/Fe] vs. [Fe/H] trend shows an enhancement of $\sim$+0.15 dex in the thick disk trend at [Fe/H]  $\lesssim$ $-$0.4. The trend gradually decreases with increasing metallicity reaching [V/Fe] $\sim$+0.1 dex at [Fe/H] $\sim$$-$0.1. The mean $\langle$[V/Fe]$\rangle$ ratio of the thick disk is +0.14 dex and $\langle\sigma\rangle$ = 0.04 dex. The thin disk [V/Fe] ratio is relatively constant and nearly zero apart from a slight increase at supersolar metallicities. The average thin disk $\langle$[V/Fe]$\rangle$ ratio is $\sim$0 dex and $\langle\sigma\rangle$ = 0.05 dex.

The ratio of [Cr/Fe] in the disk components exhibit flat trends throughout the whole metallicity range apart from slight enhancements at the highest metallicities in each Galactic component. The mean $\langle$[Cr/Fe]$\rangle$ ratio is $\sim$0 dex with average $\langle\sigma\rangle \sim$0.04 dex for the thin and thick disks.

The lowest [Mn/Fe] ratio is observed in the thick disk trend with the running mean reaching down to $\sim$$-$0.3 dex at [Fe/H] $\sim$$-$0.6. The trend steadily increases with increasing metallicity and attains [Mn/Fe] $\sim$$-$0.1 dex at [Fe/H] $\sim$$-$0.1. The mean  $\langle$[Mn/Fe]$\rangle$ ratio in the thick disk is $\sim$$-$0.2 dex and $\langle\sigma\rangle$ = 0.08 dex. The lowest [Mn/Fe] value of the thin disk is $\sim$$-$0.2 dex at [Fe/H] $\sim$$-$0.4. The trend also increases steadily to [Mn/Fe] $\sim$$-$0.02 dex at [Fe/H] $\sim$+0.1. The average  $\langle$[Mn/Fe]$\rangle$ ratio in the thin disk is $-$0.14 dex with $\langle\sigma\rangle$ = 0.06 dex.

The thick disk trend of cobalt gradually decreases from [Co/Fe] $\sim$+0.2 dex at [Fe/H] $\sim$$-$0.7 to [Co/Fe] $\sim$+0.1 dex at [Fe/H] $\sim$$-$0.1. The thin disk trend goes down from [Co/Fe] $\sim$+0.1 dex at [Fe/H] $\sim$$-$0.4 to [Co/Fe] $\sim$+0.05 dex at [Fe/H] $\sim$$-$0.1. At  [Fe/H] $\gtrsim$$-$0.1, the thin disk trend  starts to increase up to [Co/Fe] $\sim$+0.1 dex at supersolar metallicities. The average $\langle$[Co/Fe]$\rangle$ for our thick disk trend is +0.16 dex with $\langle\sigma\rangle$ = 0.04 dex and for the thin disk $\langle$[Co/Fe]$\rangle$ = +0.06 dex with $\langle\sigma\rangle$ = 0.04 dex, thus showing a hint of a separation.

The separation is also less distinct in the [Ni/Fe] vs. [Fe/H] trend: the thick disk trend is rather flat with $\langle$[Ni/Fe]$\rangle$ = 0.05 dex and $\langle\sigma\rangle$ = 0.03 dex; the thin disk trend shows a slight elevation at supersolar metallicites, but the running mean remains relatively flat with $\langle$[Ni/Fe]$\rangle$ about 0 dex and $\langle\sigma\rangle$ = 0.03 dex.

\subsection{Galactic Bulge}\label{gb_res}
\begin{figure*}[!th]
\centering  
\includegraphics[width=\linewidth]{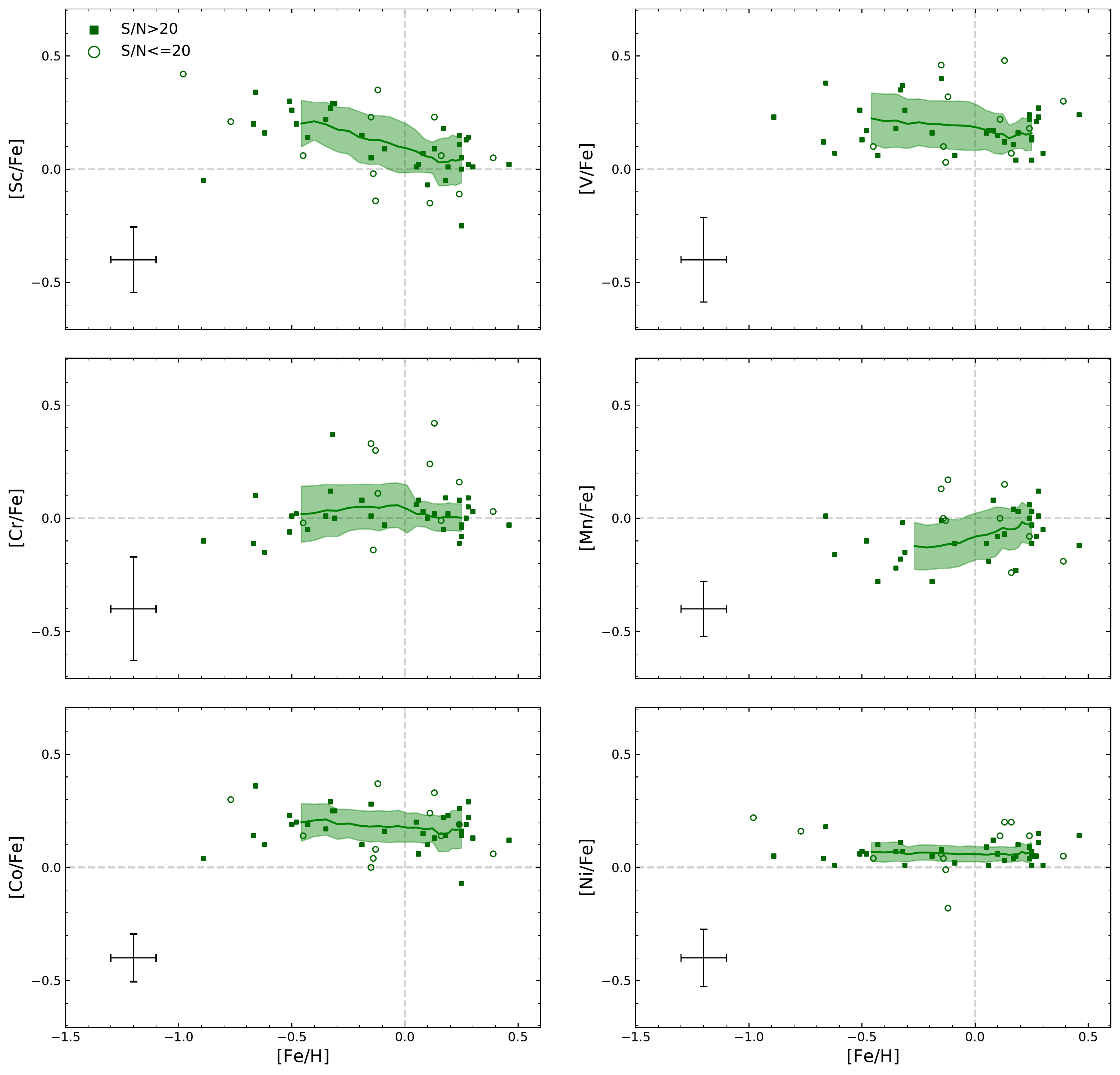} 
\caption[Galactic Bulge abundance trends]{[X/Fe] vs. [Fe/H] trends (LTE) of the examined Bulge giants: green squares denote stars with spectra having an S/N ratio above 20, while the green empty circles denote stars having spectra with an S/N ratio $\leq$ 20. The solid lines represent the running means and shaded areas show the 1$\sigma$ scatter (only for stars with S/N ratio > 20). The typical uncertainties for each elements from Table \ref{mc_lbl} are also shown. In the plots we use A(Sc)$_\odot$ = 3.04 \citep{peh17_sc2}, A(V)$_\odot$ = 3.89, A(Cr)$_\odot$ = 5.62, A(Mn)$_\odot$ = 5.42,  A(Fe)$_\odot$ = 7.47, A(Co)$_\odot$ = 4.93, A(Ni)$_\odot$ =  6.20 \citep{scott15}. Note that for spectra with S/N $\leq$ 20 the uncertainty in [X/Fe] becomes significantly large \citep[see][]{jonsson17_d}.}
\label{bulge_plt}\end{figure*}

The Bulge sample consists of 45 K giants in five different fields: 11 stars in the SW field, 10 in B3, 8 in BW, 11 in B6, and 5 in BL. \cite{jonsson17_d} concluded that a S/N ratio below 20 has a very strong negative impact on the precision and accuracy of the determined stellar parameters and abundances. For this reason, we only used the abundances obtained from stellar spectra with S/N ratio above 20 (about 30 stars) when calculating the running mean and 1$\sigma$ scatter. The stars with the S/N ratio below 20 are, however, still plotted in Figure \ref{bulge_plt} but marked differently.

In the Bulge we observe one decreasing (Sc) and one increasing (Mn) trend with higher metallicity. The remaining trends of V, Cr, Co, and Ni do not change significantly as a function of [Fe/H] in the Bulge.

The [Sc/Fe] ratio decreases from $\sim$+0.2 dex at [Fe/H] $\sim$$-$0.5 to $\sim$+0.05 dex at [Fe/H] $\sim$+0.15, where it becomes flat. The average $\langle$[Sc/Fe]$\rangle$ ratio  is +0.1 dex and $\langle\sigma\rangle$ = 0.10 dex.
The [V/Fe]  vs. [Fe/H] trend is somewhat decreasing with higher metallicity with $\langle$[V/Fe]$\rangle \sim$+0.18 dex and $\langle\sigma\rangle$ = 0.09 dex.
The [Cr/Fe] ratio of $\sim$+0.03 dex at [Fe/H] $\lesssim$ 0 is slightly enhanced compared to [Cr/Fe] $\sim$0 dex at supersolar metallicities. The 1$\sigma$ spread at [Fe/H] $\lesssim$ 0 is also significantly larger. On average the $\langle$[Cr/Fe]$\rangle$ ratio is 0.03 dex and $\langle\sigma\rangle$ = 0.09 dex.
The [Mn/Fe] ratio is steadily increasing from [Mn/Fe] $\sim$$-$0.15 dex at [Fe/H] $\sim$$-$0.3 to [Mn/Fe] $\sim$0 dex at [Fe/H] $\sim$+0.25.  The mean $\langle$[Mn/Fe]$\rangle$ ratio is $-$0.07 dex and $\langle\sigma\rangle$ = 0.1 dex. 
The trend of Co is more or less constant  with $\langle$[Co/Fe]$\rangle$ = +0.17 dex and $\langle\sigma\rangle$ = 0.07 dex. The [Ni/Fe] ratio does not change significantly with metallicity either, showing $\langle$[Ni/Fe]$\rangle$ = +0.06 dex and $\langle\sigma\rangle$ = 0.04 dex.

\subsection{Uncertainties in the determined abundances}
\subsubsection{Systematic uncertainties}\label{sec:syst_uncert}
Generally, the origin of systematic errors lies in incorrectly determined stellar parameters, model atmosphere assumptions and/or atomic data, and are often quite difficult to estimate.

Abundances of the iron-peak elements examined in this work have been measured for the Gaia benchmark stars. These are well studied stars with careful stellar parameter and abundance determinations using several different methodologies \citep{jofre15}; three of these stars overlap with our sample: $\alpha$Boo, $\beta$Gem and $\mu$Leo. A discussion about the quality of our stellar parameters derived from the FIES spectra and other works including the  benchmark values can be found in \cite{jonsson17_d}. Briefly, our parameters were found to fall within the uncertainties of the Gaia benchmark results, except for $\log g$ of $\mu$Leo which is slightly higher.

 The comparison between our determined abundances and the Gaia benchmark values are shown in Table \ref{benchm_abund}. Our abundance results fall within the uncertainties of the benchmark values for V and Ni. For Sc, only the abundance of $\alpha$Boo falls within the uncertainties of the Gaia benchmark values. However, remember that we used the VALD line list for Sc that accounts for hfs instead of the Gaia-ESO list which does not. For Cr, our results for $\mu$Leo and $\alpha$Boo fall within the uncertainties of the benchmark values, while the results for $\beta$Gem, do not. Regarding Co, the result for $\mu$Leo falls within the uncertainties of the benchmark values.

The largest discrepancies are observed for Mn: 0.17, 0.34 and 0.22 dex for $\beta$Gem, $\mu$Leo  and $\alpha$Boo respectively. In the analysis, we had only one satisfactory Mn line which lowers the precision of the measurements. Nevertheless, if these differences are applied to our [Mn/Fe] vs. [Fe/H] trend in Figure \ref{disk_plt}, it would become significantly lower. For example, $\mu$Leo, which is located at the outermost metallicity end of the Mn trend and lies at ([Fe/H], [Mn/Fe]) = (0.23, 0.08), would appear outside of the trend as low as at (0.23, $-$0.26). Additionally, the benchmark values do not agree with the [Mn/Fe] vs. [Fe/H] trend of F and G dwarfs from \cite{bb} which is presented in Section \ref{comp_sn}.

\begin{table*}[!th]
\centering
\caption[Absolute iron-peak abundances of the overlapping Gaia benchmark stars]{Iron-peak abundances of the overlapping Gaia benchmark stars. Top row for each star: the abundances presented in \cite{jofre15}; middle row: our results; bottom row: the difference between our results and the literature values.}
\label{benchm_abund}
\begin{tabular}{cccccccc}
\hline\hline
Star & A(Sc) & A(V) & A(Cr) & A(Mn) & A(Co) & A(Ni) &  \\
  & [dex] &  [dex]  &  [dex]  &  [dex] &  [dex] &  [dex] \\\hline

$\beta$Gem & 3.28 $\pm$0.12 & 3.99 $\pm$ 0.16 & 5.66 $\pm$0.04 & 5.14 $\pm$ 0.12 & 4.91 $\pm$ 0.05 & 6.26 $\pm$ 0.05 &  \\
HIP37826 & 3.09	 & 3.92 & 5.73 & 5.31 & 4.98 & 6.24 &  \\
 & $-$0.19  & $-$0.07 &  0.07 & 0.17 & 0.07 & $-$0.02 \\\hline
$\mu$Leo & 3.45 $\pm$0.06 & 4.23 $\pm$ 0.06 & 5.91 $\pm$ 0.08 & 5.39 $\pm$ 0.20 & 5.34 $\pm$ 0.09 & 6.50 $\pm$ 0.12 &  \\
HIP48455 & 3.34 & 4.20 & 5.89 & 5.73 & 5.35 & 6.46 &  \\
& $-$0.11 & $-$0.03 & $-$0.02 & 0.34 & 0.01 & $-$0.04 &  \\\hline
$\alpha$Boo & 2.79 $\pm$ 0.14 & 3.49 $\pm$ 0.10 & 5.00 $\pm$ 0.07 & 4.41 $\pm$ 0.14 & 4.48 $\pm$ 0.05 & 5.69 $\pm$ 0.08 &  \\
HIP69673 & 2.66 & 3.41 & 4.95 & 4.63 & 4.56 & 5.64 &  \\
 & $-$0.13 & $-$0.08 & $-$0.05 & 0.22 & 0.08 & $-$0.05 & 
\end{tabular}
\tablefoot{A(X) =  $\log \left( \frac{\textnormal{N$_\textnormal{X}$}}{\textnormal{N$_\textnormal{H}$}} \right)_{*}$ = $ \left[ \frac{\textnormal{X}}{\textnormal{H}} \right] _{*} + \log \left( \frac{\textnormal{N$_\textnormal{X}$}}{\textnormal{N$_\textnormal{H}$}} \right)_{\odot}$}
\end{table*}
%3.09  3.34  2.66, diff: -0.192 -0.11  -0.132
% old values sc
% 3.20   3.40  2.66, diff: -0.08  -0.05, -0.13

\subsubsection{Random uncertainties}

Ideally, random uncertainties should be estimated for every star, but in this case, it would be very time consuming. Instead, we followed the approach of selecting a typical star in our sample which allows to assign a typical random uncertainty for each abundance trend. Although fast, this method obviously cannot reflect the true uncertainty for every individual star. E.g.,  stellar properties retrieved using spectral lines of metal poor stars are less affected by blends but suffer more from spectral noise than lines in the spectrum of metal rich stars. Nevertheless, this method is still able to give an idea about the expected size of uncertainties in the chemical abundances.  
\begin{itemize}
\item To estimate how random uncertainties in the stellar parameters affect the measured abundances, we chose $\alpha$Boo to be the typical star. Our observed FIES spectrum has a high S/N ratio, allowing us to isolate random uncertainties originating more or less solely from the parameters. We generated a set of normally distributed random uncertainties  with a standard deviation of 50 K for $T_{\textnormal{eff}}$, 0.15 dex for $\log g$, 0.05 dex for [Fe/H] and 0.1 km/s for $v_{\textnormal{mic}}$ for the disk sample. In total, 500 synthetic sets of the stellar parameters were created. These uncertainties were then added to the stellar parameters used in the original abundance measurements for $\alpha$Boo. The same procedure was repeated for the Bulge sample, again using the stellar parameters of $\alpha$Boo, but the standard deviations were assumed to be twice as large compared to the disk. 

The standard deviation obtained from the spread in abundances from the analysis of the synthetic spectra is denoted as $\sigma_{\textnormal{param}}$ in Table \ref{mc_lbl} and shows the uncertainty in the abundances due to random uncertainties in the stellar parameters. Note that this method assumes that the uncertainties are uncorrelated producing an overestimated value. The results of the simulation for the disk and Bulge are shown in Figure \ref{mc_disk_hist} and \ref{mc_bulge_hist}.

\item To estimate other sources of uncertainties we also calculated the line-by-line dispersion, $\sigma_{\textnormal{lbl}}$,  of the abundances for each element apart from Mn, for which only one line was analysed. For the disk sample we again chose the FIES spectrum of $\alpha$Boo as the representative star, whereas for the Bulge, the spectrum of B3-f1 was used as this star has typical stellar parameters and a typical S/N ratio of $\sim$30 while the S/N ratio of the $\alpha$Boo spectrum is too high to represent the Bulge sample. The line-by-line scatter represents a combined uncertainty originating from the continuum placement, S/N ratio, uncertainties in $\log gf$-values, unknown line blends, and shortcomings of the model atmospheres \citep[e.g.,][]{johnson14}. For Mn, the line-by-line dispersion was assumed to be the mean of the values calculated for the other elements. This, of course, does not include the uncertainties originating from the atomic data for the \ion{Mn}{I} line used. Note that the disk spectra from \cite{thygesen12} have much lower S/N ratios and, consequently,  larger $\sigma_{\textnormal{lbl}}$ than $\alpha$Boo. The spectra of $\alpha$Boo and B3-f1 from which the line-by-line abundance scatter was obtained are presented in Figure \ref{arc_spec} and \ref{b3f_spec} respectively; the $\sigma_{\textnormal{lbl}}$ values can be found in Table \ref{mc_lbl}. 

\item The formula for the total uncertainty, $\sigma_{\textnormal{total}}$, was adapted from \cite{mik17}:
\begin{equation}
\sigma_{\textnormal{total}} = \sqrt{\sigma_{\textnormal{param}}^2+\left(\frac{\sigma_{\textnormal{lbl}}}{\sqrt{N}}\right)^2}
\label{sigma_tot}\end{equation}
where $\sigma_{\textnormal{param}}$ is the uncertainty due to the stellar parameters, $\sigma_{\textnormal{lbl}}$ is the line-by-line dispersion and $N$ is the number of lines used in the analysis for each element.
 
The typical uncertainties obtained in this way are shown  in Table \ref{mc_lbl} and were used as the final uncertainty estimations. 
\end{itemize}

\begin{table*}[h]
\centering
\caption[Typical abundance uncertainties due to the stellar parameters and line-by-line abundance scatter]{Typical abundance uncertainties for the disk and Bulge stellar samples. $\sigma_{\textnormal{param}}$ represents the uncertainty due to the changes in the stellar parameters using the FIES spectrum of $\alpha$Boo; $\sigma_{\textnormal{lbl}}$ represents the uncertainty derived from the line-by-line abundance scatter for each element using the FIES spectrum of $\alpha$Boo (disk) and B3-f1 (Bulge); $\sigma_{\textnormal{total}}$ represents the combined uncertainty as in Eq. \ref{sigma_tot}.}
\label{mc_lbl}
\begin{tabular}{lccccccc}
\hline\hline
Uncertainty & Sc & V & Cr &  Mn &  Co  &   Ni  &  Component  \\ \hline
$\sigma_{\textnormal{param}}$ [dex]  &      0.07  & 0.1       & 0.08  &  0.03  & 0.05  &  0.06 &Disk  \\
                            &   0.1		&	 0.2   	& 0.1     &  0.06	   &0.1    	 & 0.1  			&Bulge\\\hline
$\sigma_{\textnormal{lbl}}$ [dex] & 0.01 	    & 0.08		& 0.09	 	& 0.06 	     & 0.05	 	& 0.05 		&Disk  \\
					   & 0.04	    & 0.08       & 0.2       &0.1         & 0.06      & 0.1     &Bulge\\\hline
%				  	   & 0.10 		& 0.10 		& 0.22 		& 0.10 		& 0.04 		& 0.10		& Bulge \\\hline
$\sigma_{\textnormal{total}}$ [dex] &0.07	 	& 0.1		& 0.1 	&  0.07		&  0.06		&  0.06		&Disk\\
  						 & 0.2 &   0.2 &   0.2 &  0.1 &  0.1 &  0.1     &Bulge\\                       
                             %&0.16 & 0.19 & 0.22 &  0.13 & 0.10 & 0.12&Bulge\\

\end{tabular}
\end{table*}

\subsection{NLTE investigation}\label{nlte_fe_peak}

The LTE assumption might not be valid in the outer atmospheric layers of giant stars. Therefore, we discuss how non-LTE (NLTE) corrections might alter the observed abundance trends of the iron-peak elements investigated.  A departure from LTE can affect a specific line's  opacity, source function, and/or the ionisation rates, and therefore the ionisation balance, in the latter case mostly affecting the minority species. Due to the different internal structures of giants and dwarfs, NLTE effects are likely to impact spectral lines of these stellar types differently, leading to an additional source of a systematic offset, from which giant-dwarf comparisons can suffer. Again, this stresses the importance of a giant-giant comparison, which reduces the offset in abundance, in the given context.

The two elements for which we were able to calculate NLTE corrections are Mn and Co. The corrections were taken from \cite{berg_mn} for Mn and \cite{berg_co} for Co\footnote{ Available online at \texttt{nlte.mpia.de}}. We note that the corrections from \cite{berg_mn} do not account for the hyperfine splitting (hfs) of \ion{Mn}{I} lines.

\begin{figure*}[h]
\centering  
\includegraphics[width=\linewidth]{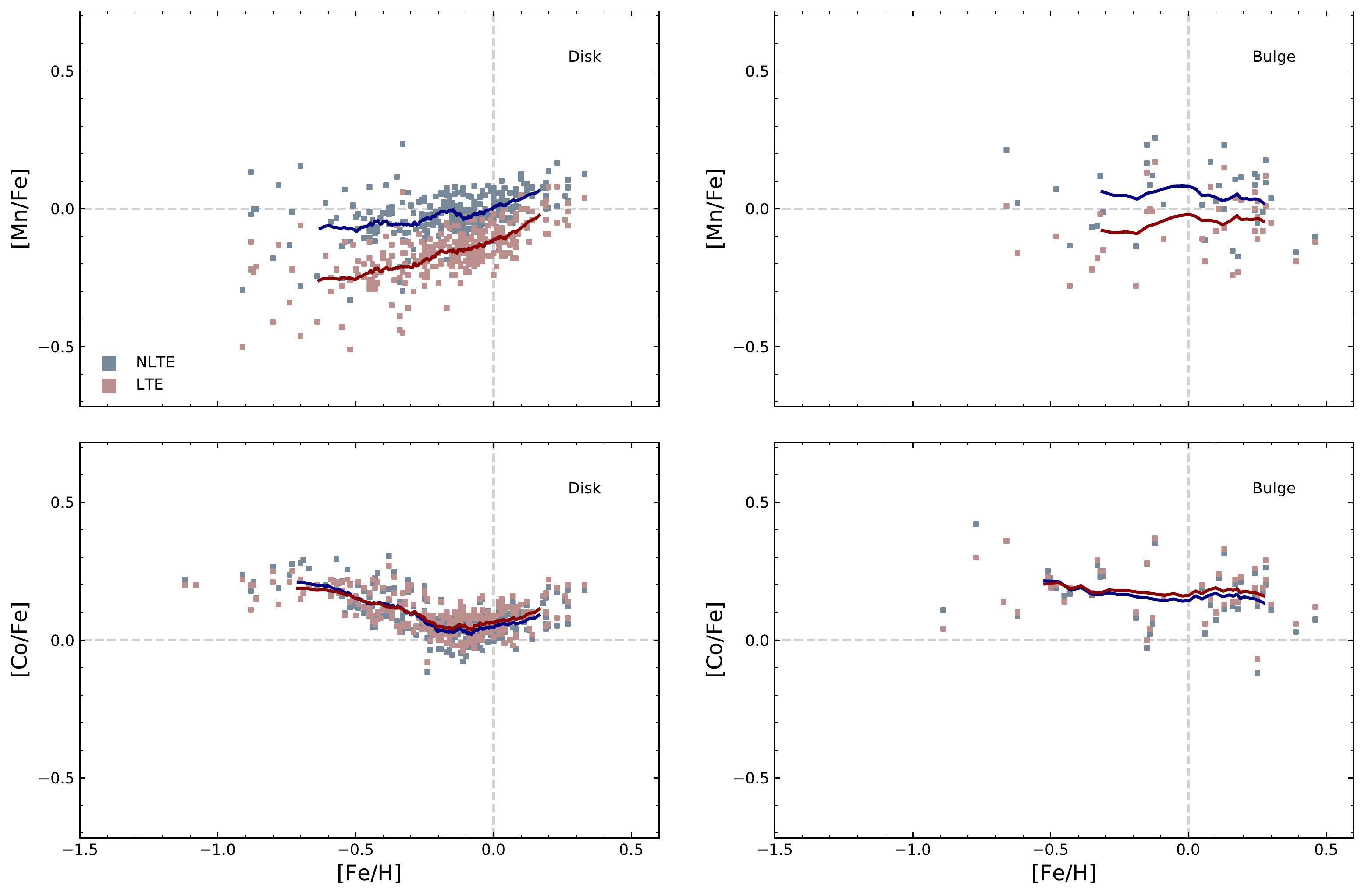}
\caption{Abundances of Mn and Co calculated for the LTE (pink) and NLTE (blue) cases in the disk (left panels; the thin and thick disks combined) and Bulge (right panels). The solid lines represent the running means (LTE: red, NLTE: dark blue). Note, that the LTE/NLTE running mean of the Bulge sample was calculated regardless of the S/N ratio, therefore the LTE running means are different from the ones in Figure \ref{bulge_plt}.}
\label{nlte_co_trend}\end{figure*}

In Figure \ref{nlte_co_trend}, we plot NLTE and LTE abundances and running means for our two samples. Overall, the differences for the [Co/Fe] ratios are not large, and as a result, the trends remain relatively unchanged. For the trend of Mn, on the other hand, the difference is prominent and the [Mn/Fe] ratio increases drastically, especially at low [Fe/H], and slightly flattens out in the disk and Bulge.

For the remaining elements, we were not able to calculate NLTE corrections directly, however, in some cases, literature studies can give an idea about what one could expect.

In the analysis, we examined lines corresponding to \ion{Sc}{ii} which is by far the majority species in the Sun and K giants in the given temperature range ($\sim$3900 - 4900 K).  Indeed, \ion{Sc}{ii} is not as sensitive to NLTE effects as \ion{Sc}{i} lines. \cite{zhang08} calculated NLTE corrections for their solar \ion{Sc}{ii} values and the differences were shown to be small ($\Delta$[Sc/Fe] = $-$0.03 dex). Unfortunately, no information about NLTE corrections for giants were presented.

Previous studies of metal-poor dwarfs and giants have shown a discrepancy between abundance trends obtained from \ion{Cr}{i} and \ion{Cr}{ii} lines:  \ion{Cr}{i},a minority species, has a decreasing trend with decreasing metallicity compared to a rather flat trend of \ion{Cr}{ii}, a majority species, \citep[e.g.,][]{johnson02, lai08}. \cite{berg_cr} found that this discrepancy disappears when NLTE effects are taken into account.  \ion{Cr}{i} is also a minority species in our K giants.  The difference in the magnitude of NLTE corrections between the metal-rich and metal-poor stars for \ion{Cr}{i} might be $\sim$0.05 dex given the results in \cite{jofre15}. 
 % meaning it is more sensitive to NLTE effects than the majority species, \ion{Cr}{ii}. 
In the Gaia benchmark sample, the NLTE corrections for \ion{Cr}{i} were calculated for the three overlapping stars \citep{jofre15}. For the metal-poor $\alpha$Boo ([Fe/H]= $-$0.57), the NLTE abundance differs by +0.09 dex from the LTE case, whereas for the more metal-rich $\beta$Gem ([Fe/H] = 0.08)  and $\mu$Leo ([Fe/H] = 0.20) the corrections are +0.06 and +0.05 dex respectively.

NLTE corrections for V are still not available, as discussed in \cite{scott15, bb, jofre15}. For our K giants, \ion{V}{i} is a minority species and might, therefore, be sensitive to NLTE effects. We are, unfortunately, unable to predict the magnitude of the NLTE corrections. %If it is also subjected to over-excitation and over-ionisation like some other neutral iron-peak elements, then NLTE corrections would increase the most with decreasing metallicity \citep{bergemann14}.

Similarly to V, there are no extensive works on NLTE corrections for Ni lines \citep{jofre15}. However, \cite{scott15} argue that NLTE corrections for \ion{Ni}{i} are probably small for the Sun in the optical region. In the case of cooler giants, \ion{Ni}{i} is most probably the dominant species having a relatively large ionisation potential. %, and NLTE effects should be small. 

%We expect that our results for \ion{V}{i} and \ion{Cr}{i} are more likely to suffer from NLTE effects since they are the minority species in the examined K giants. 
 %For a comparative analysis as this work, this should not be an issue since the disk and Bulge K-giants have similar atmospheric properties, hence, the magnitude of NLTE corrections are expected to be similar.

%__________________________________________________________________
\section{Discussion}

In this section, we discuss the disk-Bulge abundance trends of Sc, V, Cr, Mn, Co and Ni and compare them to some literature studies. Again we stress that this is a homogeneous giant-giant analysis of stars in the solar neighbourhood and the Bulge. 
\subsection{Disk and Bulge: comparison}\label{d_b_comp}

\begin{figure*}[!th]
\centering  
\includegraphics[width=\linewidth]{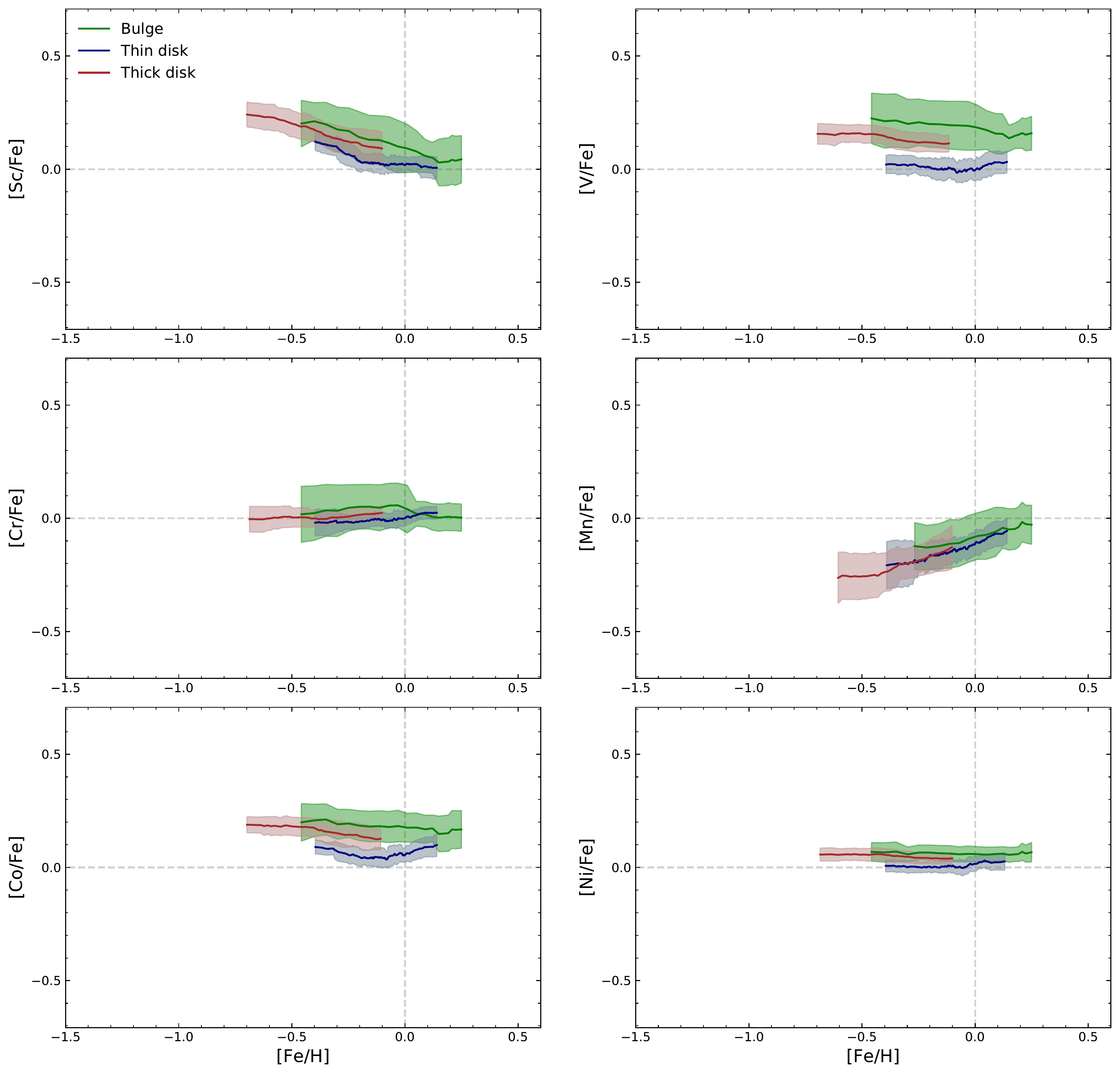} 
\caption[Solar neighbourhood and Galactic Bulge: comparison (running means)]{The running means and 1$\sigma$ scatter of the thick disk (red), thin disk (blue) and Bulge (green) trends (same as in Figure \ref{disk_plt} and \ref{bulge_plt}).}
\label{disk_bulge_plt}\end{figure*}

To see more clearly how the trends from the local thin and thick disks and the Bulge relate to each other, we plot them together in Figure \ref{disk_bulge_plt}. In this plot we include the running means and running 1$\sigma$ scatter of the trends in the disk and Bulge from Figure \ref{disk_plt} and \ref{bulge_plt} respectively. These trends will be more extensively discussed below, but to summarize, the Bulge trends tend to be more enhanced than the disk trends, although with a generally high 1$\sigma$ scatter. The enhancement is largest in the trend of [V/Fe] vs [Fe/H], but it might also be enhanced in the trend of Co, although the trends' uncertainty-bands mostly overlap.

\subsubsection{Scandium}

Scandium (here and hereafter we refer to stable isotopes only) is mostly produced in Ne burning or through the radioactive progenitor $^{45}$Ti in explosive Si and O burning (\citealt{ww95}, hereafter WW95). It has a complex formation background, being predominantly produced in SNe II, similarly to $\alpha$-elements \citep[e.g.,][]{bb}, and having a dependence on the properties of the progenitor stars such as metallicity and mass (e.g., WW95; Figure 5 in \citealt{nomoto13}  and references therein). These relations should make Sc sensitive to the environment in which it is produced.  

As shown in Figure \ref{disk_bulge_plt}, in our [Sc/Fe] vs. [Fe/H] thick disk trend, there is a hint that [Sc/Fe] might be slightly more enhanced than the thin disk trend at comparable metallicities, although this is within the uncertainties. The enhancement of the thick disk in Sc has also been observed in the studies of dwarf stars in \cite{reddy03,reddy06, bb} as well as \cite{adibek12} up to solar metallicity. The Bulge Sc trend shows a somewhat enhanced running mean than the thick disk, and at larger [Fe/H], where only the thin disk is present, the Bulge trend remains slightly enhanced w.r.t. the thin disk. Note, however, that the 1$\sigma$ scatter between the three trends is very prominent and non-negligible.

%The relation between [Sc/Fe] and [Fe/H] can reflect the SNe II/SNe Ia ratio in a Galactic component. Among the three galactic regions, our thin disk has the lowest [Sc/Fe] ratio, which is consistent with the assumption that the thin disk stars are generally younger \citep[e.g.,][]{bens14_d}, i.e., they formed when the yield contribution from SNe Ia was already significant. Hence, the opposite is valid for the thick disk: many of those stars are old and were formed before SNe Ia started to occur. In the Bulge, the majority of the stars are old as well, i.e., $\gtrsim$10 Gyr \citep[e.g.,][]{clarkson08, valenti13, gennaro15}, and the enhanced Bulge trend could be attributed to its higher metallicity, and potentially, a higher SFR. 

Present day theoretical models in e.g., \cite{koba11} cannot reproduce observed Sc trends, and the formation of Sc is still poorly understood. In order to draw any definite conclusions from the abundance trends of Sc, we need to better understand how it is produced. Potentially, multidimensional 2D \citep[e.g.,][]{maeda03, tominaga09} and 3D \citep{janka12} models, or the $\nu$-process in some environments \citep{iwamoto06, heger10, koba11} may improve the theoretical yields in the future.

\subsubsection{Vanadium}
Vanadium is predominantly produced in explosive Si and O burning (WW95). SNe Ia are thought to create more V than SNe II do (C03), but, in fact, little is known about the production of V. Vanadium yields produced in nucleosynthesis models do not show a strong metallicity dependence, but as in the case of Sc, they reproduce deficient [V/Fe] ratios \citep[e.g.,][]{koba11}.

Our V trend for the thick disk is enhanced w.r.t. the thin disk by approximately +0.1 dex on average, see Figure \ref{disk_bulge_plt}. A similar enhancement of the thick disk is also observed in e.g., \cite{reddy03,reddy06,adibek12, bb}. The mean [V/Fe] ratio of the Bulge trend is quite similar to the thick disk: the average difference between them is only +0.05 dex. But the trends in Figure \ref{disk_bulge_plt} do not look very similar. 
%The overall disk trend of V resembles somewhat the Sc disk trend generally decreasing with increasing [Fe/H]. The Bulge trend of V, on the contrary, appears to be much flatter than the Sc trend in the Bulge.

It is rather difficult to say what the V trend tells us about the evolution and formation of the disk and the Bulge since this element is not well-explored. From our trend, we can conclude that [V/Fe] shows an enhancement in the Bulge compared to the thick and thin disk. However, a significant overlap between the 1$\sigma$ scatter of the Bulge and thick disk trends is present, and it increases with decreasing metallicity.

\subsubsection{Chromium}\label{disc_cr}

Similarly to V, Cr is predominantly made in explosive Si and O burning (WW95). The largest yields of Cr are thought to be created in SNe Ia (C03). However, a non-negligible amount of Cr yields is also produced in SNe II, and given their higher frequency, the overall amount of Cr created in SNe Ia and SNe II is comparable (C03).  According to some nucleosynthesis computations, Cr does not show any strong variability in SNe II yields and the amount produced in SNe Ia is very similar to the amount of Fe \citep{kobayashi11}, resulting in a flat theoretical [Cr/Fe] vs. [Fe/H] trend. As shown in Figure \ref{disk_bulge_plt}, the overall disk trend of Cr does agree with the theoretical predictions apart from a possible, slight increase in [Cr/Fe] at supersolar metallicities. This elevation of the (thin) disk trend, if real, has not been observed in, e.g., \cite{reddy03} and \cite{bens14_d}. We note that the \ion{Cr}{i} NLTE corrections for overlapping Gaia benchmark stars in Section \ref{nlte_fe_peak}, although they cannot be applied directly here, show a small difference between the metal-poor and metal-rich ends of the trend ($\Delta=0.05$ dex). Another cause for a possible elevation may be any potentially poorly modelled blends that increase with increasing metallicity. However, we have taken care to try to avoid using blended lines, see Section \ref{sec:linedata}.

The Bulge trend, on the contrary, is non-flat and enhanced at [Fe/H] $\lesssim$ 0 dex, but it decreases towards [Cr/Fe] $\sim$ 0 dex at [Fe/H] $\gtrsim$ 0 dex. This enhancement can be attributed to the larger scatter in the Cr abundance trend as shown in Figure \ref{bulge_plt}. The overlap between the Bulge and disk components is very large. 

Due to a large 1$\sigma$ overlap between the disk and Bulge trends that are, on average, roughly flat, we conclude that Cr is likely to be insensitive to the formation environment. %Similar flat Cr trends have been observed for the disk dwarfs and Bulge giants in \cite{johnson14} as well as for the Bulge and disk dwarfs in \cite{bens17}. 

\subsubsection{Manganese}

Manganese is created in explosive Si burning and $\alpha$-rich freeze-out (C03). The amount of Mn yields depend, however, on the properties of the progenitor white dwarf. Computations have shown that SNe Ia events occurring at the Chandrasekhar-mass produce more Mn than Fe, independent of the metallicity of the white dwarf \citep[e.g.,][]{nomoto97, yama15}. Interestingly, SNe Ia events taking place below the Chandrasekhar-mass underproduce Mn instead, but the amount of Mn yields increases with metallicity \citep[e.g.,][]{woo11}. Early nucleosynthesis models indicated metallicity dependent Mn yields produced in SNe II which showed to be very large in comparison to the yields from SNe Ia (e.g., WW95). However, more recent simulations of necleosynthesis in SNe II produce Mn yields that are smaller at all metallicities and cannot explain the increasing [Mn/Fe] vs. [Fe/H] trend without considering the SN Ia contributions \citep{koba06,sukhbold16}. Therefore, the main production source of Mn is presumed to be SNe Ia which makes it suitable for probing the SNe Ia/SNe II ratio in different systems. 

Our thin and thick disk trends for Mn in Figure \ref{disk_bulge_plt} strongly overlap each other and the running means of the two trends have roughly the same slope at all metallicities. The Bulge trend is marginally enhanced compared to the disk components and has a very similar slope. At the lowest metallicities, our Bulge trend starts to increase, which we believe is an artificial effect due to a larger scatter. Also, as the equivalent width of the rather weak \ion{Mn}{I} line used decreases, it becomes more sensitive to perturbations such as spectral noise, blends, etc., lowering the precision of the measurements. As a result, the dispersion of the thick disk trend increases for [Fe/H] $\lesssim$ -0.3. Overall, the three trends appears to be very similar.

Many studies on Mn have been carried out, and some of them have found a different behaviour of the thin and thick disks. For example, \cite{feltz07} examined disk dwarfs and concluded that  the thick disk stars have a steadily increasing [Mn/Fe] ratio with increasing [Fe/H], whereas the thin disk stars have a flat trend up and until [Fe/H] $\sim$ 0 dex and an increasing trend thereafter. However, the [Mn/Fe] vs. [Fe/H] trend of dwarfs in the solar neighbourhood presented in \cite{bb}  shows a separation that is in agreement with ours:  an increasing trend with increasing [Fe/H] both in the thin and thick disk (in LTE). Similar thin and thick disk trends are also presented in \cite{reddy03, reddy06, adibek12}. 

Regarding the Bulge, various studies have shown an agreement between Mn abundance trends in the Bulge and the overall disk trend including the Bulge giants from \cite{barb13} and the disk dwarfs from \cite{reddy03,reddy06} (see Figure 11 in \cite{mcw16} and references therein).

Based on the observed increasing [Mn/Fe] ratios with increasing metallicity, \cite{gratton89} suggested that Mn might be overproduced in SNe Ia compared to Fe.  \cite{mcw16} argues that given a mere overproduction of Mn in SNe Ia, one could expect deficient [Mn/Fe] ratios in $\alpha$-rich systems where the contribution from SNe II has been large, e.g., in the Bulge. The trends of Mn in the thick disk would then also be deficient compared to the thin disk, which has only been seen in \cite{bb} when NLTE corrections from \cite{berg_mn} were applied resulting in a relatively flat overall disk trend. However, while the nucleosynthesis models in \cite{kobayashi11} and \cite{nomoto13} can reproduce the observed LTE trends of Mn rather well, they are not able to explain the flat NLTE trend. This suggests that the NLTE corrections might not be correct (as we discussed in Section \ref{nlte_fe_peak}, the hfs was not taken into account when calculating the corrections) or/and the  models might not be complete. In any case, LTE abundances of Mn, as in this work, suggest similar enrichment rates of Mn in the disk and the Bulge.

%$^{55}$Mn is thought to be overproduced in SNe Ia compared to Fe. Observations have supported this theory, exhibiting a plateau in the [Mn/Fe] vs. [Fe/H] trend at [Fe/H] $\lesssim$ -1 dex which changes to a steadily increasing [Mn/Fe] ratio with increasing metallicity \citep[e.g.,][]{gratton89}. It takes about 10 Gyr for the contributions from SNe Ia to start occurring due to longer lifetimes of low-mass stars explaining the rise of the manganese produced compared to iron. 
\subsubsection{Cobalt}

Cobalt is mainly created in explosive Si burning and the $\alpha$-rich freeze-out through the radioactive progenitor $^{59}$Cu as well as by the s-process (C03).  According to the nuclesynthesis model in \cite{kobayashi11} and \cite{koba11}, Co produces a flat trend having similar SNe yields as Cr, which is not supported by the observations.  As a possible solution, \cite{kobayashi11} suggest that hypernovae\footnote{ Hypernovae are very energetic (by a factor of 10 more than for a regular SN II) core-collapse supernovae with masses M $\geq$ 20 M$_\odot$ \citep{koba11}.} (HNe) can solve the issue since they increase Co yields.  \cite{mcw16} argues, however, that HNe, apart from producing higher Co abundances, will also result in an underabundant [Cr/Fe] ratio which has not been observed.

The [Co/Fe] trend in the thick disk is enhanced by $\sim$+0.1 dex compared to the thin disk, which generally agrees with the findings in \cite{reddy03,reddy06, adibek12, bb}. In the Bulge the [Co/Fe] ratio is higher than in the thick disk at comparable metallicities with a strong 1$\sigma$ overlap between the trends. If significant, this enhancement would suggest that the thick disk and the Bulge would have experienced different chemical enrichment paths. \cite{johnson14} also note a larger  [Co/Fe] ratio in the Bulge than in both disk components.

\subsubsection{Nickel}
Nickel is a product of explosive Si burning and $\alpha$-rich freeze-out.
SNe Ia give the highest Ni yields, but SNe II, being more frequent, result in a comparable total production of Ni (C03). This element is known to produce a tight [Ni/Fe] vs. [Fe/H] trend since many clean Ni lines are available in the optical region for various stellar types \citep[e.g.,][]{jofre15}.

Our thick disk trend is enhanced in Ni compared to the thin disk by $\sim$+0.05 dex, which generally agrees with the findings in \cite{reddy03,reddy06,adibek12}. The enhancement is quite small and it matches the overall enhancement of the Bulge. If true, it could be explained assuming that the amount of SNe II nucleosynthesis products is higher in the thick disk and Bulge than in the thin disk due to, e.g., a higher SFR. There is also a significant overlap between the 1$\sigma$ scatter of the thick disk and Bulge trends. \cite{bens17} observe a similar Ni trend in the Bulge which falls on top of the thick disk trend. \cite{johnson14}, on the contrary, find an enhanced Ni trend in the Bulge compared to the thick disk at [Fe/H] $\gtrsim$ $-$0.4, which we do not see.

\subsection{ Detailed Comparison with Selected Literature Trends}\label{comp}
\subsubsection{Solar-Neighbourhood Trends by \cite{bens14_d} and \cite{bb}}\label{comp_sn}

\begin{figure*}[h]
\centering  
\includegraphics[width=\linewidth]{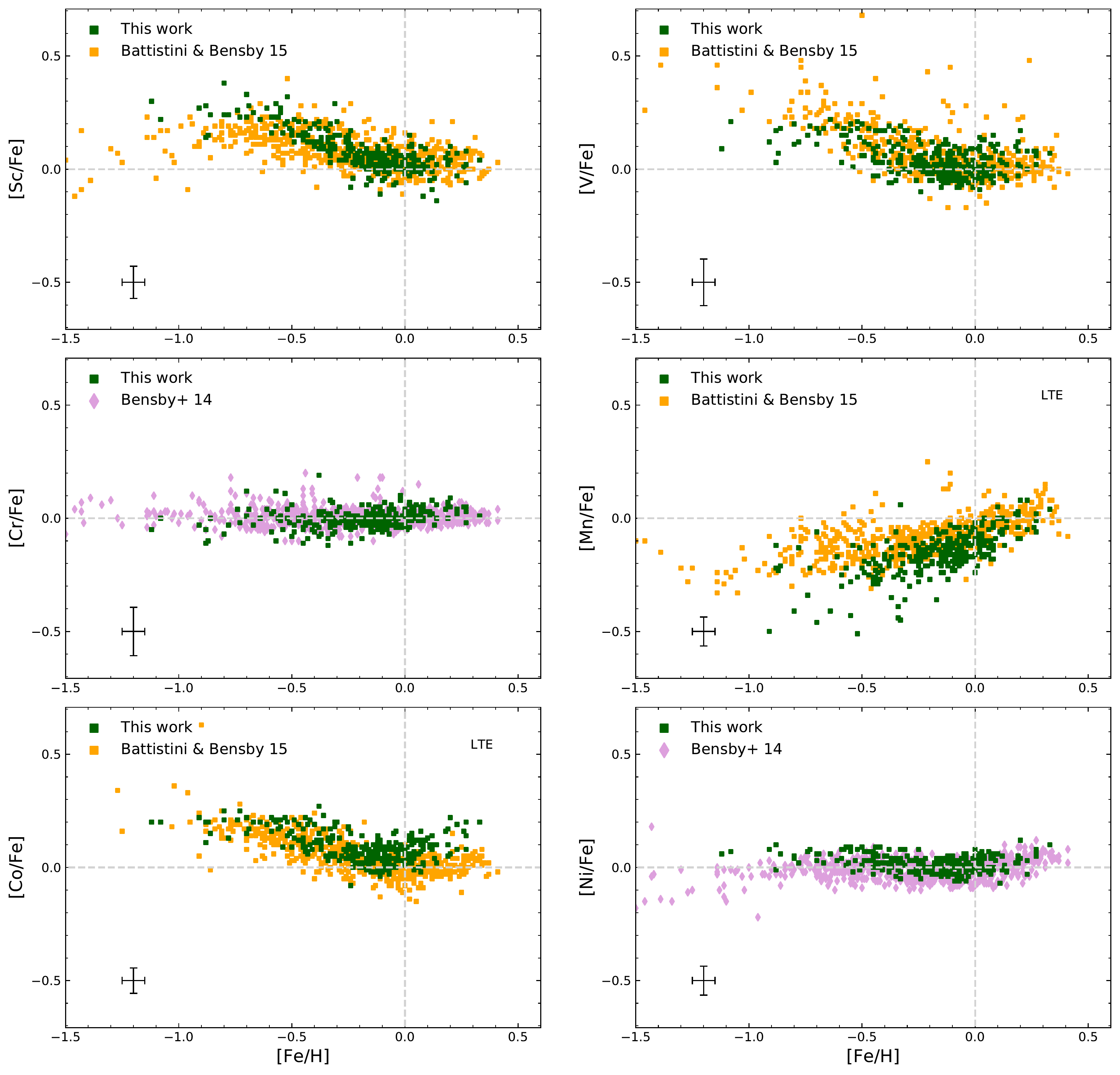} 
\caption[Disk trends: comparison]{[X/Fe] vs. [Fe/H] trends (LTE) determined for the disk giants in this work (green) together with the dwarf disk trends from \cite{bb} (orange) and \cite{bens14_d} (purple). The typical uncertainties for our stars are shown as in Figure \ref{disk_plt}.}
\label{disk_comp}\end{figure*}

\begin{figure*}[h]
\centering  
\includegraphics[width=\linewidth]{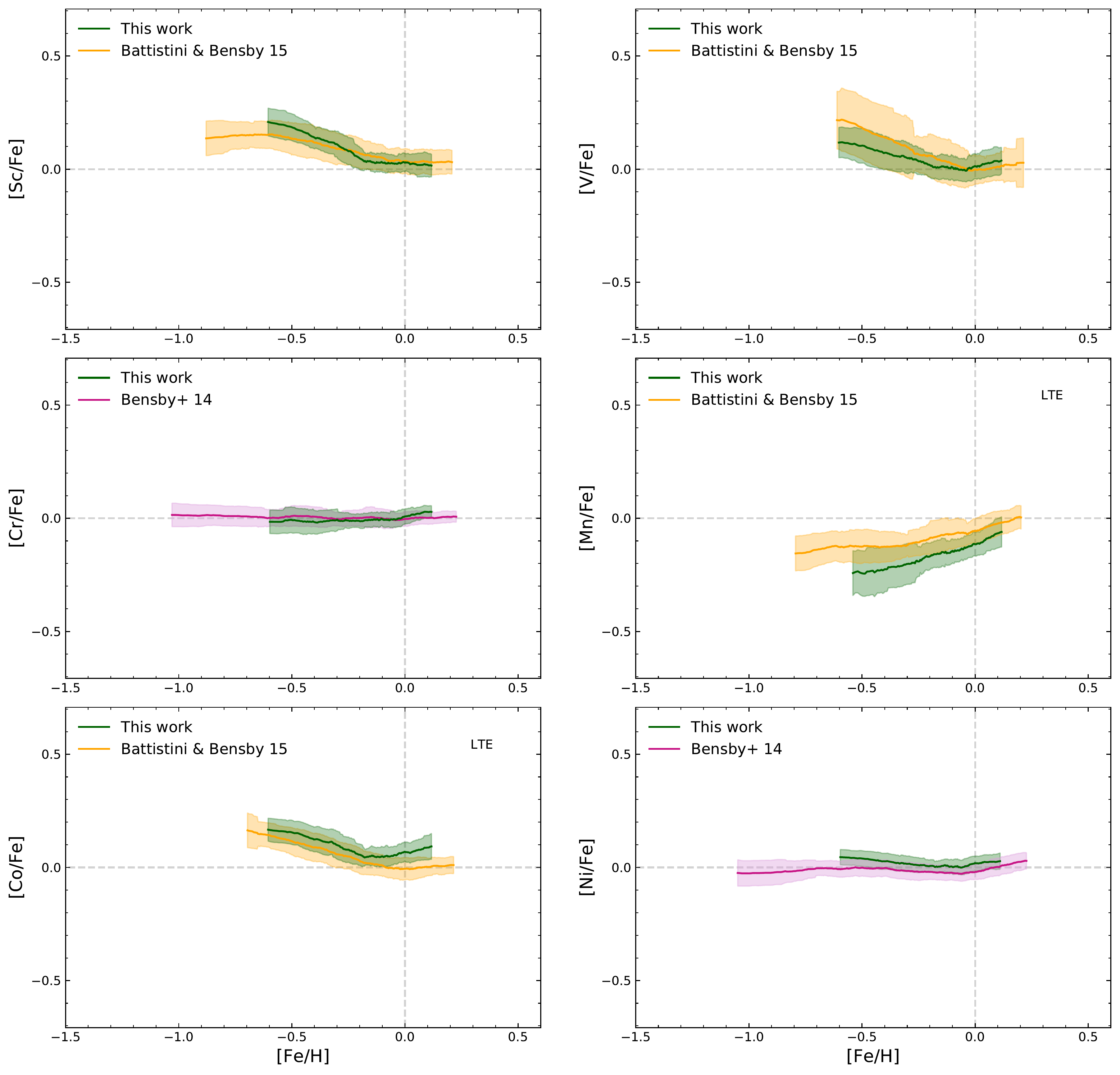} 
\caption[Disk trends: comparison (running means)]{Running means and 1$\sigma$ scatter for the disk samples (thin and thick combined) in Figure \ref{disk_comp}: this work (green),  \cite{bb} (orange) and \cite{bens14_d} (purple).}
\label{disk_comp_rm}\end{figure*}

%We did not find any published results on iron-peak elements in giants in the solar neighbourhood similar to this work. Those available contain abundances determined through an automated pipeline for a large sample of stars. Therefore, in order to check the validity of our results,
 In this section, we compare our results to the analysis of 714 F and G dwarf and sub-giant stars from the thin and thick disk in  \cite{bens14_d} (R = 40 000$-$110 000, S/N = 150$-$300, $\lambda$ = $\sim$3600$-$9300 \r{A}) and \cite{bb} (R = 45 000$-$120 000, S/N = 150$-$300, $\lambda$ = $\sim$3600$-$9300 \r{A}.
The comparison is shown in Figure \ref{disk_comp}-\ref{disk_comp_rm}. In general, the results from our giants match the results from the dwarfs, both regarding trends and scatter, but with two obvious exceptions: our [Mn/Fe] vs. [Fe/H] trend is steeper, and we have an upturn in [Co/Fe] for the highest metallicities. Note that due to the strictly differential analysis of the dwarfs, we do not adapt the dwarf trends to our solar values from \cite{scott15} and \cite{peh17_sc2} in Figure \ref{disk_comp}-\ref{disk_comp_rm}.

For Sc, our [Sc/Fe] values follow the dwarf trend at higher metallicities down to [Fe/H] $\sim$$-$0.2 dex. At [Fe/H] $\lesssim$ $-$0.2 our Sc trend starts to increase more rapidly with decreasing metallicity than that of the dwarfs (see Figure \ref{disk_comp_rm}) and follows the upper envelope of the dwarf [Sc/Fe] ratio, as shown in Figure \ref{disk_comp}. It is difficult to say what has caused this difference. NLTE effects are not expected to strongly affect \ion{Sc}{ii} lines as it is the majority species in K giants and FGK dwarfs.

Our V trend, on the other hand, shows an underabundance at [Fe/H] $\lesssim$ $-$0.3 compared to the dwarfs, as shown in Figure \ref{disk_comp_rm}. At [Fe/H] $\gtrsim 0$ , our [V/Fe] ratio becomes overabundant instead.
%, and the whole trend seems to be slightly shifted compared to the dwarf values. It is worth mentioning that due to the strictly differential analysis of the dwarfs, our results and the literature trends in Figure \ref{disk_comp} and \ref{disk_comp_rm} may not be scaled w.r.t. the same solar abundances, which might cause this shift.
This could potentially be due to molecular and/or atomic blends affecting our lines, which would give an metallicity dependent offset. However, it seems unlikely, since we observe the same trend in all four \ion{V}{i} lines used.

Our [Cr/Fe] ratio shows a similar flat feature as in \cite{bens14_d} with roughly the same spread, apart from the previously discussed, possible slight increase at supersolar metallicities.

Compared to our Mn trend (LTE),  the study of dwarf stars by \cite{bb} stretches out to lower metallicities,  where it seems to flatten out  (see Figure \ref{disk_comp}). This behaviour is expected assuming a lower contribution of Mn from SNe II dominating at earlier times \citep[e.g.,][]{koba11}. Observation of extremely metal-poor halo stars ($-$4.0 < [Fe/H] < $-$2.7 ) in \cite{cayrel04} also exhibit a similar flat trend; however, in a study of halo stars by \cite{honda04} ($-$3.1 < [Fe/H] < $-$2.4), the [Mn/Fe] vs. [Fe/H] trend is significantly scattered.

This is not the only difference between our Mn (LTE) trend and the literature. At [Fe/H] $>0$, the discrepancy is smaller, but as the metallicity decreases, our trend decreases at a higher rate following the lower envelope of the dwarf trend. In the analysis, we used only one \ion{Mn}{i} line (four lines are used in \cite{bb}) which lowers the overall precision. Also, note that if the Gaia benchmark values from Section \ref{sec:syst_uncert} were adapted here, the discrepancy at lower [Fe/H] would increase even more, e.g., the metal-poor $\alpha$Boo would appear at ($-$0.57, $-$0.44) instead of ($-$0.57, $-$0.22). 

\begin{figure*}[h]
\centering  
\includegraphics[width=\linewidth]{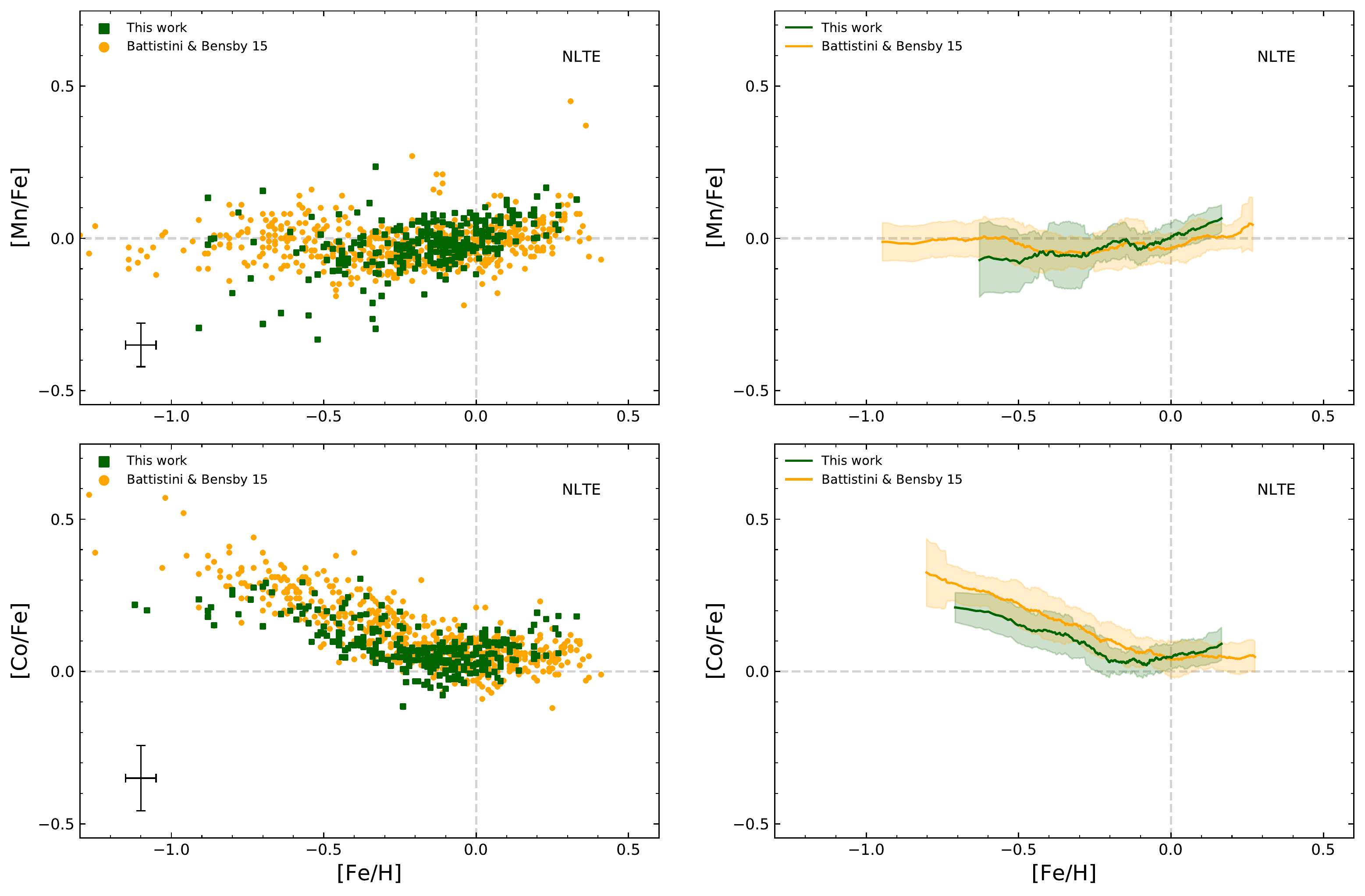}
\caption[]{Left panels: the NLTE Mn and Co trends for our disk giants (green) and dwarfs from \citep{bb} (orange). Right panels: the running means and 1$\sigma$ scatter for the trends in the left panel (same colours). The typical uncertainties for our stars are shown as in Figure \ref{disk_plt}.}
\label{nlte_bat}\end{figure*}

When NLTE corrections are applied, the difference becomes less severe, as shown in the top row in Figure \ref{nlte_bat}. Nevertheless, our NLTE trend does not flatten out as much, although the decrease in the [Mn/Fe] ratio at low [Fe/H] could again be attributed to the larger scatter.

In \cite{bb}, the [Co/Fe] vs. [Fe/H] trend resembles the trend for Sc, having a plateau at supersolar metallicities followed by an increase with decreasing [Fe/H] which flattens out for [Fe/H] $\lesssim$ $-$0.5.  Up to the solar values, our [Co/Fe] vs. [Fe/H] trend follows the literature values, although it is slightly enhanced, which could potentially be connected to the choice of the solar Co abundance. As for the [V/Fe] vs. [Fe/H] trend but more prominently, our [Co/Fe] values show a significant increase up to [Co/Fe] $\sim$+0.1 dex at the supersolar metallicities. Again, one of the explanations could be atomic or molecular blends, the risk of which increases with higher [Fe/H]. However, all of the four \ion{Co}{i} lines used in the analysis show the same increase, hence, again, line blending is unlikely to explain this upturn.

Since NLTE corrections were available for our giants and also the dwarfs in \cite{bb}, we can check if NLTE effects can be the reason for the discrepancy. In the left bottom panel in Figure \ref{nlte_bat}, we plot the NLTE Co abundances and in the right bottom panel we show the running mean and 1$\sigma$ scatter of the trends. As for the LTE trend of V, the trends seem to be shifted, in a metallicity-dependent way, which might be a sign of blending. However, the NLTE-corrections applied increases the match between our and the reference trends, possibly indicating that the corrections applied are too small for the highest metallicities and too large for the lowest. In conclusion, we do not fully understand the origin of this divergence, and the deviation might come from the model atmospheres, which is difficult to assess.

The [Ni/Fe] vs. [Fe/H] trends also show an agreement between our values and those published in \cite{bens14_d}. Our trend is somewhat enhanced as shown in Figure \ref{disk_comp_rm}. This may simply be attributed to the solar Ni abundance used. Furthermore, the spreads of both trends are similar for the metallicity range of our sample.

\subsubsection{ Galactic-Bulge Trends} \label{b_comp}

\begin{figure*}[!th]
\centering  
\includegraphics[width=\linewidth]{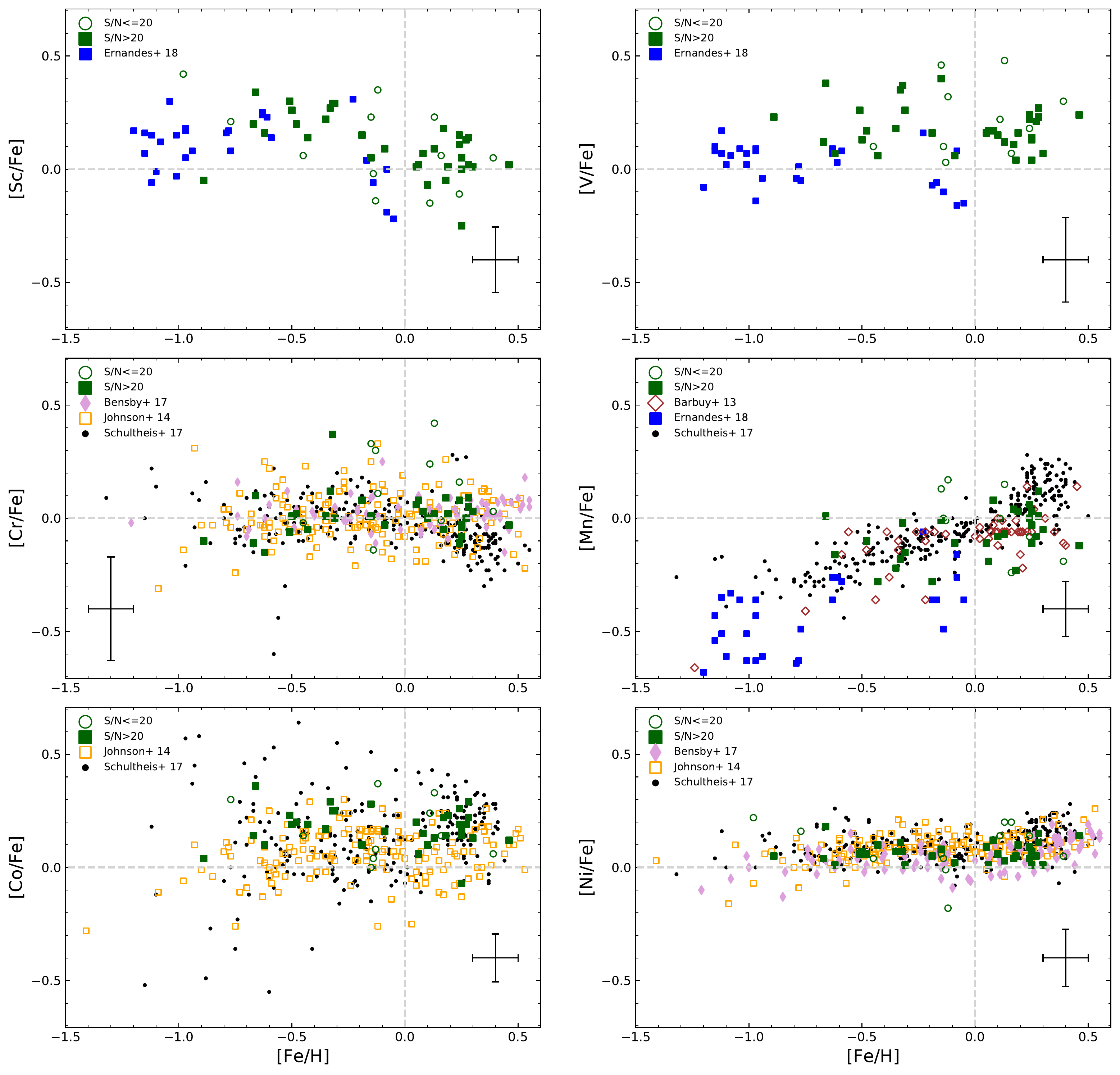} 
\caption[Bulge trends: comparison]{[X/Fe] vs. [Fe/H] trends (LTE) determined for the Bulge giants in this work: green squares for S/N ratios above 20, green open circles otherwise. Note that for spectra with S/N < 20 the uncertainty in [X/Fe] becomes significantly large \citep[see][]{jonsson17_d}. Also plotted: Bulge giant trends from \cite{barb13} (open red diamonds), \cite{ernandes18} (blue squares), \cite{johnson14} (open orange squares) and \cite{schultheis17} (black dots) as well as the microlensed Bulge dwarfs from \cite{bens17} (purple filled diamonds). The typical uncertainties for our stars are shown as in Figure \ref{bulge_plt}.}
\label{bulge_comp}\end{figure*}

\begin{figure*}[h]
\centering  
\includegraphics[width=\linewidth]{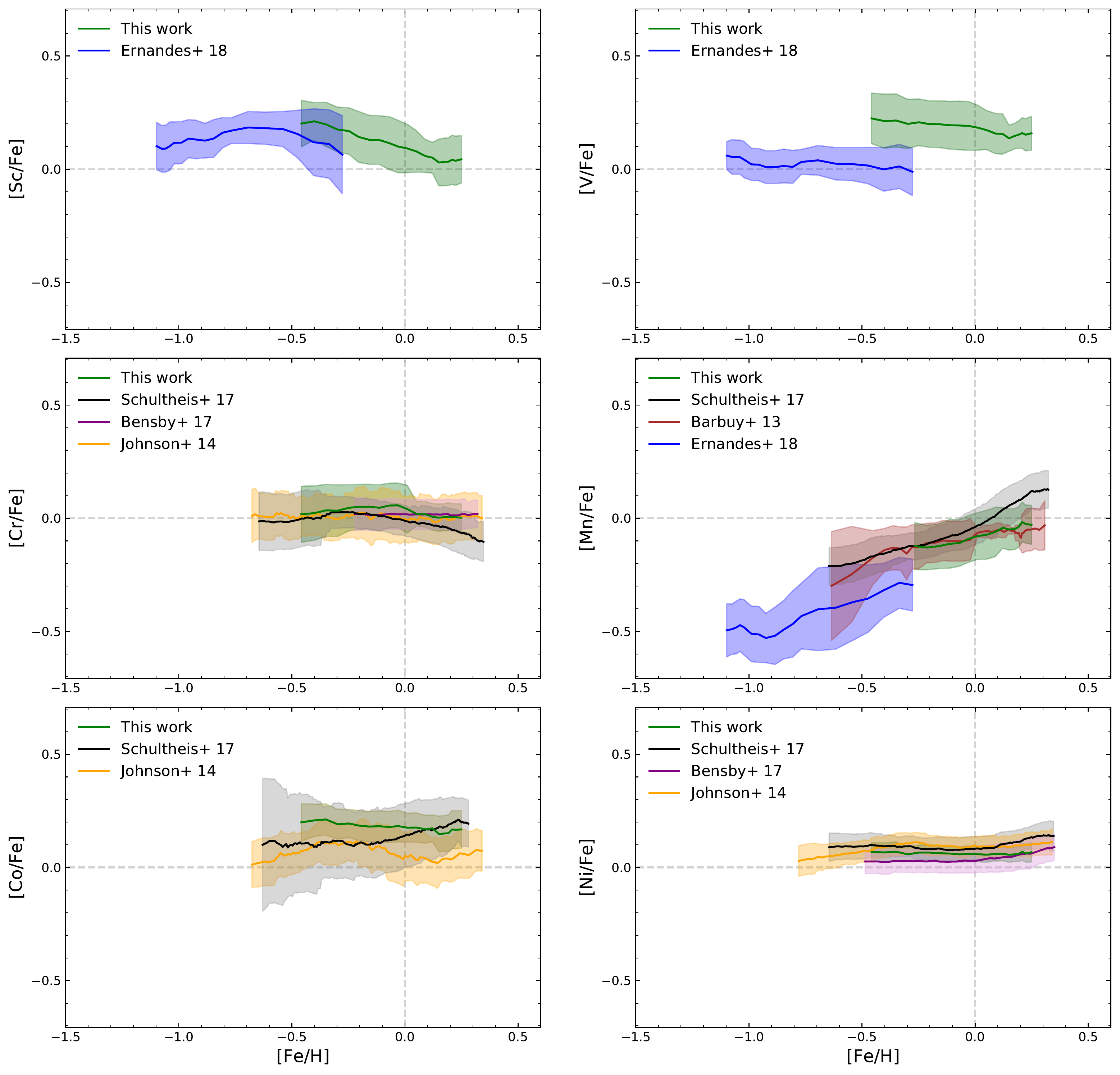} 
\caption[Bulge trends: comparison (running means)]{Running means and 1$\sigma$ scatter for the Bulge samples in Figure \ref{bulge_comp}: this work (green; only spectra with an S/N ratio > 20), \cite{barb13} (red), \cite{ernandes18} (blue), \cite{johnson14} (orange), \cite{schultheis17} (black/grey) and \cite{bens17} (purple).}
\label{bulge_comp_rm}\end{figure*}

 In Figures \ref{bulge_comp} and \ref{bulge_comp_rm} we plot the results from our investigation of the abundance trends in the Bulge and a comparison with other relevant literature studies. Figure \ref{bulge_comp} shows the actual data points of all studies and their trends. Figure \ref{bulge_comp_rm} shows the running means in order to easier visualize the trends.

We provide a thorough discussion on the comparison below, but briefly, we see that our abundance trends of the Bulge mostly show a general agreement with those from the literature, although a few off-sets will be discussed. %This gives us the confidence in our analysis techniques and relevance of the results obtained.

There are several high-resolution spectroscopic studies of red giants in the Bulge that provide abundances of iron-peak elements. \cite{ernandes18} (R = 45 000$-$55 000, S/N = 30$-$300, $\lambda$ = 4800$-$6800 \r{A}) have studied 28 red giants in five globular clusters in the Bulge and determined abundances of Sc, V and Mn in those stars. \cite{johnson14} (R = $\sim$20 000, S/N $\gtrsim$ 70, $\lambda$ = $\sim$5500$-$7000 \r{A}) measured abundances of Cr, Co and Ni in 156 red giants from the Galactic Bulge. \cite{schultheis17} (R = 22 500,  $\lambda$ = 1.5$-$1.7 $\mu$m) worked on 269 red giants from the infra-red APOGEE survey in the Baade's window (BW) and obtained abundances of Cr, Mn, Co and Ni. \citet{zasowski2018} also uses APOGEE to map the Bulge over a larger area of the sky. The general abundance ratio versus metallicity trends found in \cite{schultheis17} are very similar to the trends in \citet{zasowski2018}, but with the latter work consisting of more stars. For simplicity, we have chosen to plot only the trends of \cite{schultheis17} in Figures \ref{bulge_comp} and \ref{bulge_comp_rm}.
Another article on Mn by \cite{barb13} (R = 45 000$-$55 000, S/N = 9$-$70, $\lambda$ = 4800$-$6800 \r{A}) contains abundances of 56 red giants in the Bulge. Moreover, there are 30 stars in our Bulge sample that overlap with \cite{barb13}. Finally, \cite{bens17} (R = 40 000$-$90 000, S/N = $\sim$15$-$200, $\lambda$ = $\sim$3500$-$9500 \r{A}) published abundances of Cr and Ni in 90 F and G dwarfs, turn-off and sub-giant stars in the Bulge. The location of the aforementioned stars are shown in Figure \ref{bulge_field}, and their abundances are plotted in Figures \ref{bulge_comp} and \ref{bulge_comp_rm}.

These authors used LTE models, and we also plot our LTE Mn and Co trend. Note that no S/N ratio cuts were applied to the literature results.  Additionally, the different works, except the results from \cite{bens17} due to the differential analysis, are scaled to solar abundances used in this paper, i.e. that of \citet{scott15} for all elements except Sc, and \citet{peh17_sc2} for Sc.

Stars from the globular clusters in \cite{ernandes18} stretch down to much lower metallicities, but there is an overlap in [Fe/H] with our most metal-poor giants. For Sc, V and Mn, our results appear more enhanced compared to \cite{ernandes18} which is especially clear in Figure \ref{bulge_comp_rm}. For Mn, the trend from \cite{ernandes18} is lower than all the other literature studies. Moreover, their [V/Fe] ratios in the Bulge are comparable to ours in the thin disk. For this reason, it appears that the results in \cite{ernandes18} might suffer from a systematic offset. However, the shapes of their trends are quite similar to ours in the overlapping regions.

As mentioned above, 30 of our stars were the same as the red giants from  the work on Mn by \cite{barb13}. However, as seen in Figure \ref{bulge_comp}, even if there are some significant abundance differences when all stars are considered, the mean value of the discrepancies between the overlapping stars is only $-$0.01 dex with the standard deviation of 0.15 dex. For the overlapping stars, \cite{barb13} adopted the stellar parameters from \cite{zoc06} and \cite{lec07}, and the latter have been discussed in \cite{jonsson17_b}. They show that the stars with the parameters from \cite{jonsson17_b} are more spread out along the RGB in the HR-diagram and have a clearer increase in metallicity with decreasing effective temperature, which is expected from isochrones, than the stars with the parameters from \cite{lec07} \citep[see Figure 2 in][]{jonsson17_b}. This suggests that the parameters used here have a higher accuracy and precision. The overall shape and position of the [Mn/Fe] vs. [Fe/H] trend from this work  and \cite{barb13} are not severely affected by these differences, as seen from Figure \ref{bulge_comp} and \ref{bulge_comp_rm}. This suggests that the Mn trend is relatively insensitive to changes in the stellar parameters. Both trends continue to rise at roughly the same rate, whereas the slope of the Mn trend in \cite{schultheis17,zasowski2018} steepens drastically at supersolar metallicities.

The disk and Bulge dwarfs studied in \cite{bens14_d} and \cite{bens17} appear to have very similar Cr trends predominantly remaining at the solar value over the whole metallicity range as seen in Figure \ref{disk_comp_rm} and \ref{bulge_comp_rm}. \cite{johnson14} find a similar flat trend for their red giants. Our Cr trend seems enhanced at [Fe/H] < 0 dex, however, as discussed in Section \ref{disc_cr}, this is most probably due to a larger scatter between the individual stars. At [Fe/H] > 0 dex, our trend converges towards [Cr/Fe] $\sim$ 0 dex as well. Interestingly, the trend from APOGEE \citep{schultheis17,zasowski2018} changes from flat at [Fe/H] < 0 dex to decreasing at supersolar metallicities.

Our Co trend is noticeably flatter and more enhanced than the trends in \cite{johnson14} and \cite{schultheis17,zasowski2018}. However, the literature trends do not fully agree either: especially in Figure \ref{bulge_comp_rm} one can see that the trend in \cite{johnson14} is on average slightly lower than the trend of APOGEE, and the two trend shapes are somewhat different, too. Possibly these rather modest differences could be due to the small sample sizes in this work and in \cite{johnson14}. It would be interesting to investigate whether a larger stellar sample would change the shape of our trend.
%at [Fe/H] $\lesssim$ $-$0.4 dex, the APOGEE trend is flat, whereas the trend in \cite{johnson14} is decreasing with decreasing [Fe/H]; at [Fe/H] $\sim$$-$0.2 the trend from \cite{schultheis17} is increasing with increasing [Fe/H], whilst the tend from  \cite{johnson14} decreases. Finally, at [Fe/H] $\gtrsim$ +0.1 dex the trend from \cite{johnson14} starts to increase again. 

The trends for Ni seem to agree rather well, showing an upward feature at [Fe/H] $> 0$ dex. The Ni trends of giants are slightly enhanced by $\sim$+0.05 dex compared to the results in \cite{bens17} which could be a question of the adopted solar Ni abundance due to the fact that \cite{bens17} do a differential analysis against the Sun.

%__________________________________________________________________
\section{Conclusions}
Recent observations of the Bulge have revealed its boxy/peanut shape, cylindrical stellar rotation and a young stellar population, undoubtedly challenging the idea of its origin. Previously, the Bulge was thought to be a typical classical bulge formed through dissipation of gas or merging events, whereas now, in the light of the new discoveries, the idea of the secular evolution of the disk has gained more credibility. However, the true picture will almost certainly be more complicated, and apart from the disk stars, the Bulge might also contain a minor spheroidal component. %To clarify the situation, observational constraints are highly important, especially in the context of a homogeneous comparative analysis of the disk and Bulge.   

In this work, we provide observational constraints by measuring the abundances of six iron-peak elements (Sc, V, Cr, Mn, Co and Ni) in K giants and perform a homogeneous giant-giant comparison of stars in the solar neighbourhood and the Galactic Bulge. Iron-peak elements are produced in thermonuclear and core collapse SNe, and can probe the chemical enrichment path of Galactic components, although not much attention has been paid to these elements before. We use 291 high-resolution optical spectra mainly obtained using the FIES spectrograph at the NOT for the disk sample, and 45 spectra of Bulge stars collected using the UVES/FLAMES spectrograph at the VLT. To retrieve stellar chemical compositions from the spectra, we use 1-D, spherically-symmetric LTE MARCS model atmospheres and the spectral synthesiser SME. The separation of the thick and thin disk components was performed using the Gaussian Mixture Model clustering method. The components were identified according to the metallicity and [Ti/Fe] ratios taken from \cite{jonsson17_d} as well as the kinematic data (proper motions and radial velocities) from Gaia DR2 \citep{gaia16, gaia18}.

The measured abundance trends show that the thick disk is more enhanced in V and to a lesser degree in Co than the thin disk. The Bulge, in turn,  might be  even more enhanced in V and Co than the thick disk, although within the uncertainties.

We have not found any results in the literature comparing Sc and V abundances in the disk and Bulge, but for Co, \cite{johnson14} also observe an enhanced trend of the Bulge giants compared to the thick and thin disk dwarfs.
%The calculated NLTE corrections are quite small and do not change our trends significantly.
Our [Ni/Fe] ratio is very similar in the thick disk and Bulge, in agreement with the findings in \cite{bens17}, and show a strong overlap between the Ni abundances in these two trends. However, \cite{johnson14} find a more enhanced trend of Ni in the Bulge than in the thick disk at higher [Fe/H].  For Cr, we find very similar trends in all the investigated Galactic components roughly exhibiting solar values throughout the whole metallicity range, suggesting that Cr is not sensitive to the formation environment. This has also been found in  \cite{johnson14} and \cite{bens17}. The trends for Mn obtained here are again very similar in the disk and Bulge being steadily increasing with increasing metallicity at about the same rate. This is consistent with the results in, e.g., \cite{barb13} who compare their Bulge giants to, among others, the disk dwarfs in  \cite{reddy03,reddy06}. The applied NLTE corrections change the [Mn/Fe] vs. [Fe/H] trend drastically by enhancing it and decreasing the slope both in the disk and the Bulge.

While the trends of Sc, Cr, Mn and Ni suggest similar chemical enrichment in the Bulge and local (thick) disk, Co and especially V exhibit some differences. Theoretical predictions cannot reproduce the observed abundance trends of V and Co in the disk \citep[e.g.,][]{koba11, kobayashi11}. Without having a clear idea about the production mechanisms of these elements, it is difficult to draw any definite conclusion regarding the chemical enrichment history in the disk and Bulge. This issue has a complex nature since many other factors apart from the nucleosynthetic yields, such as, the IMF, SFR, gas flows, etc., play an important role in Galactic chemical evolution models.

Based solely on the observed abundance trends of the examined iron-peak elements, we conclude that the local thick disk and the Bulge might not have experienced the same evolutionary path. However, this does not necessarily contradict the fact that the Milky Way Bulge is likely to have emerged through dynamical instabilities of the disk. The chemical enrichment history of the \textit{local} thick disk might not be identical to the one of the thick disk region closer to the Bulge. A sample of thick disk stars lying closer to the Galactic centre or in the inner disk is needed to confirm or reject this hypothesis. 

%__________________________________________________________________
\begin{acknowledgements}
The anonymous referee is thanked for insightful comments and suggestions that improved the paper in several ways. This research has been partly supported by the Lars Hierta Memorial Foundation, and the Royal Physiographic Society in Lund through Stiftelsen Walter Gyllenbergs fond and M{\"a}rta och Erik Holmbergs donation.
H.J. acknowledges support from the Crafoord Foundation, Stiftelsen Olle Engkvist Byggm{\"a}stare, and Ruth och Nils-Erik Stenb{\"a}cks stiftelse.
This work has made use of data from the European Space Agency (ESA) mission {\it Gaia} (\url{https://www.cosmos.esa.int/gaia}), processed by the {\it Gaia} Data Processing and Analysis Consortium (DPAC, \url{https://www.cosmos.esa.int/web/gaia/dpac/consortium}). Funding for the DPAC has been provided by national institutions, in particular the institutions participating in the {\it Gaia} Multilateral Agreement.
This publication made use of the SIMBAD database, operated at CDS, Strasbourg, France, NASA's Astrophysics Data System, and the VALD database, operated at Uppsala University, the Institute of Astronomy RAS in Moscow, and the University of Vienna.
\end{acknowledgements}
%-------------------------------------------------------------------

%\bibliography{/Users/henrik/Documents/Bibliografi/papers.bib,/Users/henrik/Documents/Bibliografi/kurucz.bib,/Users/henrik/Documents/Bibliografi/sulphur.bib,/Users/henrik/Documents/Bibliografi/GESreferencesv5all.bib}
%\bibliographystyle{aa}

\bibliographystyle{aa}
\bibliography{references}
% % ##################### Online material ##################
 \begin{appendix}
 \section{Tables}
 
\begin{table*}[h]
\caption{Basic data for the observed solar neighbourhood giants. Coordinates and magnitudes are taken from the SIMBAD database, while the radial velocities are measured from the spectra. The S/N per data point is measured by the IDL-routine \texttt{der\textunderscore snr.pro}, see \href{http://www.stecf.org/software/ASTROsoft/DER\textunderscore SNR}{http://www.stecf.org/software/ASTROsoft/DER\textunderscore SNR}.}
\begin{tabular}{l l l l r r r l}
\hline
\hline
HIP/KIC/TYC & Alternative name & RA (J2000) & Dec (J2000) & $V$ & $v_{\mathrm{rad}}$ & S/N &  Source\\
        &                  & (h:m:s)    & (d:am:as)   &     & km/s\\
\hline
HIP1692 &           HD1690 & 00:21:13.32713 & $-$08:16:52.1625 &  9.18 &   18.37 &  114 & FIES-archive \\
HIP9884 &           alfAri & 02:07:10.40570 & +23:27:44.7032 &  2.01 &  $-$14.29 &   90 & PolarBase \\
HIP10085 &          HD13189 & 02:09:40.17260 & +32:18:59.1649 &  7.56 &   26.21 &  156 & FIES-archive \\
HIP12247 &            81Cet & 02:37:41.80105 & $-$03:23:46.2201 &  5.66 &    9.34 &  176 & FIES-archive \\
HIP28417 &          HD40460 & 06:00:06.03883 & +27:16:19.8614 &  6.62 &  100.64 &  121 & PolarBase \\
HIP33827 &           HR2581 & 07:01:21.41827 & +70:48:29.8674 &  5.69 &  $-$17.99 &   79 & PolarBase \\
HIP35759 &          HD57470 & 07:22:33.85798 & +29:49:27.6626 &  7.67 &  $-$30.19 &   85 & PolarBase \\
HIP37447 &           alfMon & 07:41:14.83257 & $-$09:33:04.0711 &  3.93 &   11.83 &   71 & Thygesen et al. (2012)\\
HIP37826 &           betGem & 07:45:18.94987 & +28:01:34.3160 &  1.14 &    3.83 &   90 & PolarBase \\
HIP43813 &           zetHya & 08:55:23.62614 & +05:56:44.0354 &  3.10 &   23.37 &  147 & PolarBase \\
\hline
\label{tab:basicdata_sn}
\end{tabular}
\tablefoot{This is only an excerpt of the table to show its form and content. The complete table is available in electronic form at the CDS.}
\end{table*}

\begin{table*}[h]
\caption{Basic data for the observed bulge giants. The S/N per data point is measured by the IDL-routine \texttt{der\textunderscore snr.pro}, see \href{http://www.stecf.org/software/ASTROsoft/DER\textunderscore SNR}{http://www.stecf.org/software/ASTROsoft/DER\textunderscore SNR}.}
\begin{tabular}{l c c c c}
\hline
\hline
Star$^a$ & RA (J2000) & Dec (J2000) & $V$ & S/N\\
         & (h:m:s)    & (d:am:as)   &     & \\
\hline
SW-09 & 17:59:04.533 & $-$29:10:36.53 & 16.153 & 16\\
SW-15 & 17:59:04.753 & $-$29:12:14.77 & 16.326 & 15\\
SW-17 & 17:59:08.138 & $-$29:11:20.10 & 16.388 & 11\\
SW-18 & 17:59:06.455 & $-$29:10:30.53 & 16.410 & 14\\
SW-27 & 17:59:04.457 & $-$29:10:20.67 & 16.484 & 13\\
SW-28 & 17:59:07.005 & $-$29:13:11.35 & 16.485 & 16\\
SW-33 & 17:59:03.331 & $-$29:10:25.60 & 16.549 & 14\\
SW-34 & 17:58:54.418 & $-$29:11:19.82 & 16.559 & 12\\
SW-43 & 17:59:04.059 & $-$29:13:30.26 & 16.606 & 16\\
SW-71 & 17:58:58.257 & $-$29:12:56.97 & 16.892 & 14\\
\hline
\end{tabular}
\label{tab:basicdata_bulge}
\tablefoot{This is only an excerpt of the table to show its form and content. The complete table is available in electronic form at the CDS.\\
\tablefootmark{a}{Using the same naming convention as \citet{lec07} for the B3-BW-B6-BL-stars.}
}
\end{table*}

\begin{table*}[h]
\caption{Atomic data for the spectral lines used in the analysis. %The first part list the lines used in the determination of the stellar parameters and calcium abundance, while the second part list the lines used to determine the oxygen, magnesium, and titanium abundances. 
All atomic data apart for Sc are collected by the Gaia-ESO line list group \citep{heiter15}. For Sc atomic data are taken from the VALD list \citep{vald99, vald17}. %For the three \ion{Ca}{i}-lines denoted with asterisks only the gravity-sensitive wings are used. 
The references listed are for $\log gf$. 
%In cases where several references are given for a quantity, the value listed is a mean of the reference values.
} 
\begin{tabular}{c c c c c }
\hline
\hline
Element & Wavelength & $\log gf$ & $\chi_{\mathrm{exc}}$ & Reference\\
        & (Å) (air)  &            & (eV)                  &      \\
\hline
\ion{Sc}{ii}  & 6245.6205  & $-$1.624  & 1.50695 & 1       \\
\ion{Sc}{ii}  & 6245.6290  & $-$2.364  & 1.50695 & 1       \\
\ion{Sc}{ii}  & 6245.6309  & $-$1.795  & 1.50695 & 1       \\
\ion{Sc}{ii}  & 6245.6362  & $-$3.364  & 1.50695 & 1       \\
\ion{Sc}{ii}  & 6245.6380  & $-$2.181  & 1.50695 & 1       \\
\ion{Sc}{ii}  & 6245.6396  & $-$2.002  & 1.50695 & 1       \\
\ion{Sc}{ii}  & 6245.6438  & $-$2.946  & 1.50695 & 1       \\
\ion{Sc}{ii}  & 6245.6454  & $-$2.148  & 1.50695 & 1       \\
\ion{Sc}{ii}  & 6245.6468  & $-$2.273  & 1.50695 & 1       \\
\ion{Sc}{ii}  & 6245.6499  & $-$2.712  & 1.50695 & 1       \\
\hline
\label{tab:linedata}
\end{tabular}
\tablefoot{This is only an excerpt of the table to show its form and content. The complete table is available in electronic form at the CDS.}
\tablebib{\\
(1) \citet{K09} 
(2) \citet{LD}
(3) \citet{W} 
(4) \citet{sobeck07}
(5) \citet{K10}
(6) \citet{BW} 
(7) \citet{C} 
(8) \citet{W14}
}
\end{table*}

\begin{table*}[h]
\caption{Stellar parameters and determined abundances for observed solar neighbourhood giants. [Fe/H] is listed in the scale of \citet{scott15}.}
\begin{tabular}{l c c c c c c c c c c}
\hline
\hline
HIP/KIC/TYC & $T_{\mathrm{eff}}$ & $\log g$ & [Fe/H] & $v_{\mathrm{mic}}$ & A(Sc) & A(V) & A(Cr) & A(Mn) & A(Co) & A(Ni)\\ 
\hline
HIP1692   &  4216  &  1.79  & $-$0.26  &  1.55  &  2.90  &  3.68  &  5.32  &  4.99  &   ...  &  5.98\\
HIP9884   &  4464  &  2.27  & $-$0.21  &  1.34  &  2.90  &  3.70  &  5.42  &  5.05  &   ...  &  5.98\\
HIP10085  &  4062  &  1.44  & $-$0.32  &  1.63  &  2.81  &  3.57  &  5.23  &  4.90  &   ...  &  5.90\\
HIP12247  &  4790  &  2.71  & $-$0.04  &  1.40  &  3.00  &  3.81  &  5.58  &  5.20  &   ...  &  6.14\\
HIP28417  &  4746  &  2.56  & $-$0.25  &  1.40  &  2.92  &  3.68  &  5.38  &  4.96  &   ...  &  5.97\\
HIP33827  &  4235  &  1.99  &  0.01  &  1.50  &  3.05  &  3.90  &  5.62  &  5.34  &   ...  &  6.23\\
HIP35759  &  4606  &  2.47  & $-$0.15  &  1.42  &  2.92  &  3.78  &  5.47  &  5.10  &   ...  &  6.07\\
HIP37447  &  4758  &  2.73  & $-$0.04  &  1.35  &  3.04  &  3.80  &  5.60  &  5.21  &   ...  &  6.14\\
HIP37826  &  4835  &  2.93  &  0.07  &  1.24  &  3.09  &  3.92  &  5.73  &  5.31  &  4.98  &  6.24\\
HIP43813  &  4873  &  2.62  & $-$0.07  &  1.51  &  2.98  &  3.78  &  5.59  &  5.16  &  4.83  &  6.07\\
\hline
\label{tab:abundances_sn}
\end{tabular}
\tablefoot{This is only an excerpt of the table to show its form and content. The complete table is available in electronic form at the CDS.}
\end{table*}

\begin{table*}[h]
\caption{Stellar parameters and determined abundances for observed Bulge giants. [Fe/H] is listed in the scale of \citet{scott15}.}
\begin{tabular}{l c c c c c c c c c c}
\hline
\hline
Star & $T_{\mathrm{eff}}$ & $\log g$ & [Fe/H] & $v_{\mathrm{mic}}$ & A(Sc) & A(V) & A(Cr) & A(Mn) & A(Co) & A(Ni)\\ 
\hline
SW-09  &  4095  &  1.79  & -0.15  &  1.32  &  3.12  &  4.20  &  5.80  &  5.40  &  4.78  &  6.11\\
SW-15  &  4741  &  1.96  & -0.98  &  1.62  &  2.48  &   ...  &   ...  &   ...  &   ...  &  5.44\\
SW-17  &  4245  &  2.09  &  0.24  &  1.44  &  3.17  &  4.31  &  6.02  &  5.58  &  5.36  &  6.58\\
SW-18  &  4212  &  1.67  & -0.13  &  1.49  &  2.77  &  3.79  &  5.79  &  5.28  &  4.88  &  6.06\\
SW-27  &  4423  &  2.34  &  0.11  &  1.60  &  3.00  &  4.22  &  5.97  &  5.53  &  5.28  &  6.45\\
SW-28  &  4254  &  2.36  & -0.14  &  1.44  &  2.88  &  3.85  &  5.34  &  5.28  &  4.83  &  6.10\\
SW-33  &  4580  &  2.72  &  0.16  &  1.39  &  3.26  &  4.12  &  5.77  &  5.34  &  5.23  &  6.56\\
SW-34  &  4468  &  1.75  & -0.45  &  1.63  &  2.65  &  3.54  &  5.15  &   ...  &  4.62  &  5.79\\
SW-43  &  4892  &  2.34  & -0.77  &  1.84  &  2.48  &   ...  &   ...  &   ...  &  4.46  &  5.59\\
SW-71  &  4344  &  2.66  &  0.39  &  1.31  &  3.48  &  4.58  &  6.04  &  5.62  &  5.38  &  6.64\\
\hline
\label{tab:abundances_bulge}
\end{tabular}
\tablefoot{This is only an excerpt of the table to show its form and content. The complete table is available in electronic form at the CDS.}
\end{table*}

\section{Figures}

The panels (a) and (c) in the first and forth row in Figure \ref{arc_spec} and \ref{b3f_spec} show a strongly overestimated synthetic spectrum (red) of two V I lines. This divergence is most likely the result of imprecise $\log gf$-values for these lines. In any case, these lines have not been used in the analysis, and they do not affect the fits and measurements done for the lines of interest, i.e., the adjacent Sc and Co lines.

\begin{figure*}[h]
\centering
\includegraphics[width=0.599\linewidth]{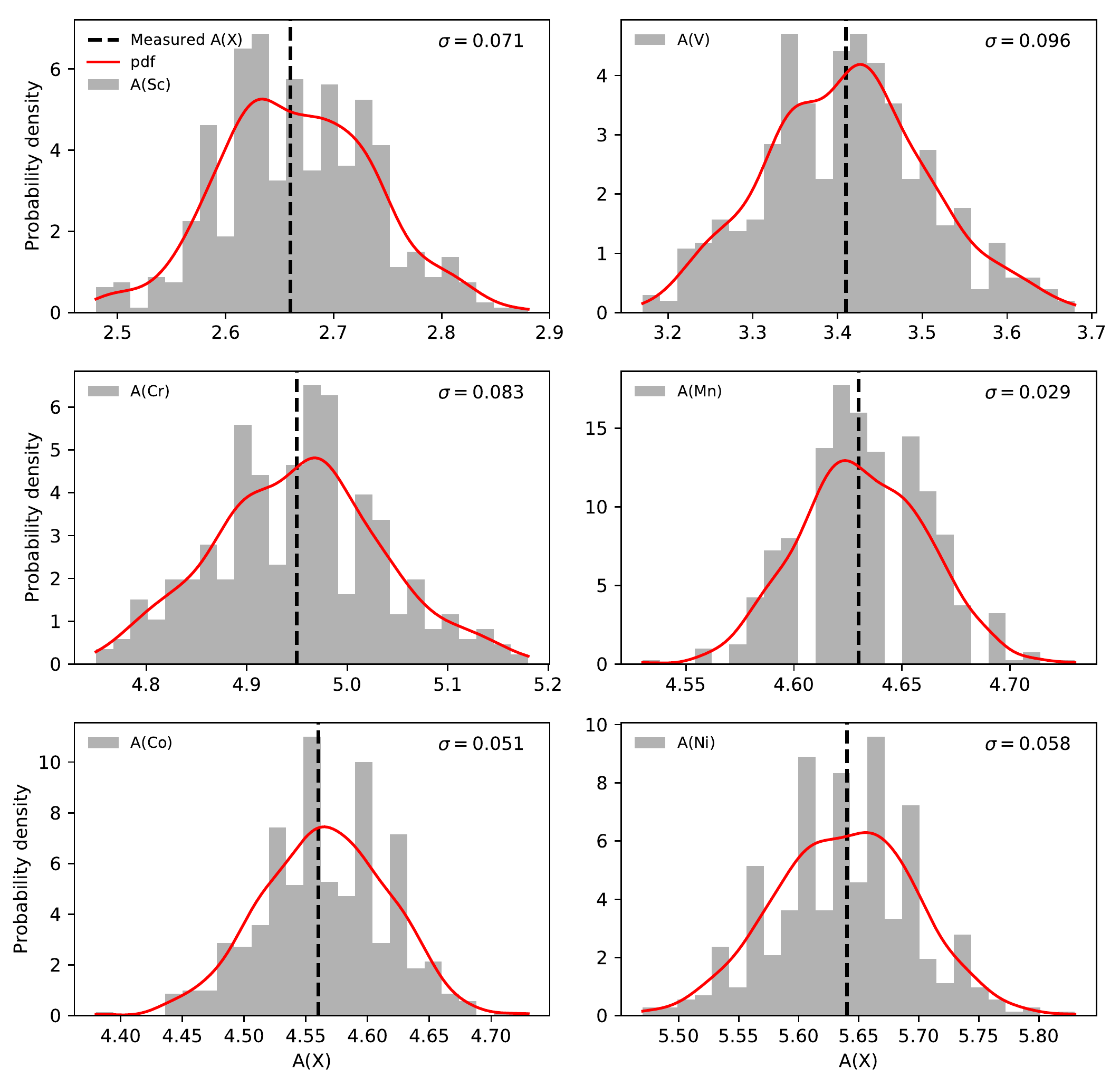}
\caption[Monte Carlo: Solar neighbourhood]{Histograms of the abundances determined from 500 synthetic data sets using the spectrum $\alpha$Boo for the disk sample. The red line denotes the estimated probability density function (pdf) for each element, and $\sigma$ is the standard deviation of each synthetic sample.} \label{mc_disk_hist}
\end{figure*}

\begin{figure*}[h]
\centering
\includegraphics[width=0.599\linewidth]{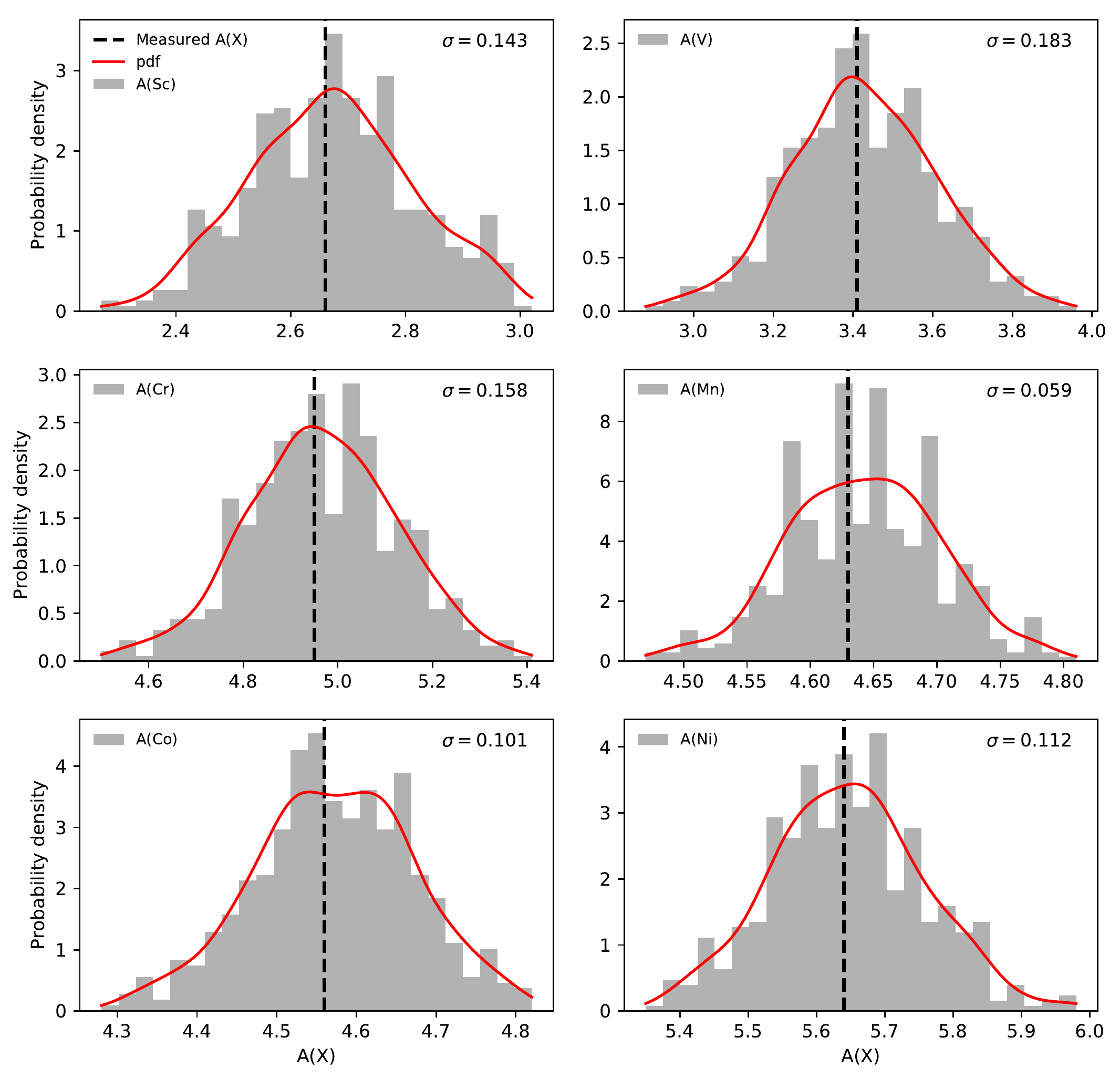}
\caption[Monte Carlo: Bulge]{Histograms of the abundances determined from 500 synthetic data sets using the spectrum $\alpha$Boo for the Bulge sample. The red line denotes the estimated probability density function (pdf) for each element, and $\sigma$ is the standard deviation of each synthetic sample.} \label{mc_bulge_hist}
\end{figure*}

\begin{figure*}[h]
\centering
\includegraphics[width=0.89\linewidth]{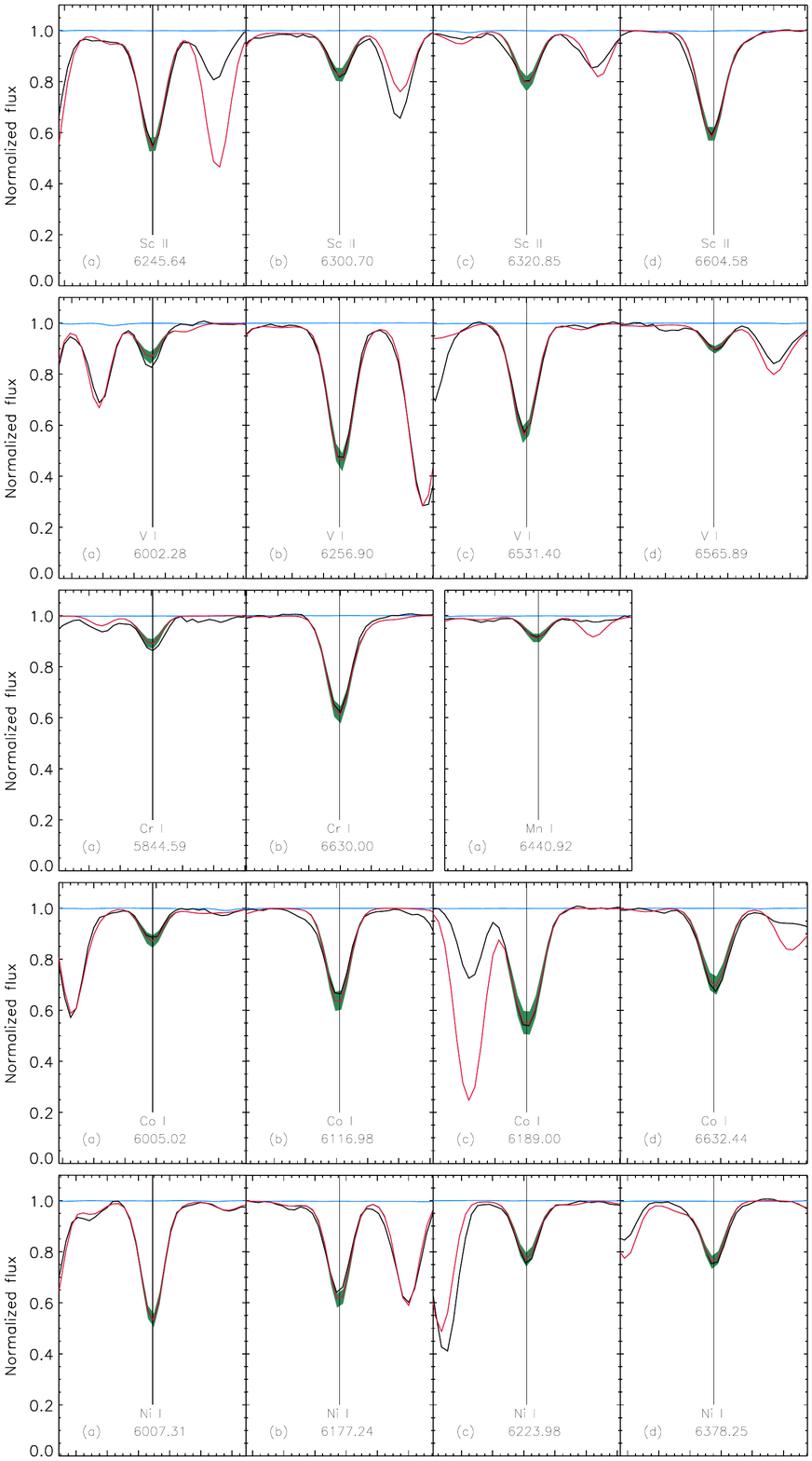}
\caption[$\alpha$Boo: spectral line fit]{Stellar lines used for the calculations of the line-by-line abundance scatter as well as the abundance determination in the analysis in the spectrum of $\alpha$Boo. The black line is the observed spectrum, the red line is the fitted spectrum and the blue line is the telluric spectrum from the $\alpha$Boo atlas of \cite{hink00} and in green is $\pm$0.2 dex of the element in question. The wavelength range of each panel is 1.2 \r{A}, i.e., the large tickmarks correspond to steps of 0.2 \r{A}.} \label{arc_spec}
\end{figure*}

\begin{figure*}[h]
\centering
\includegraphics[width=0.9\linewidth]{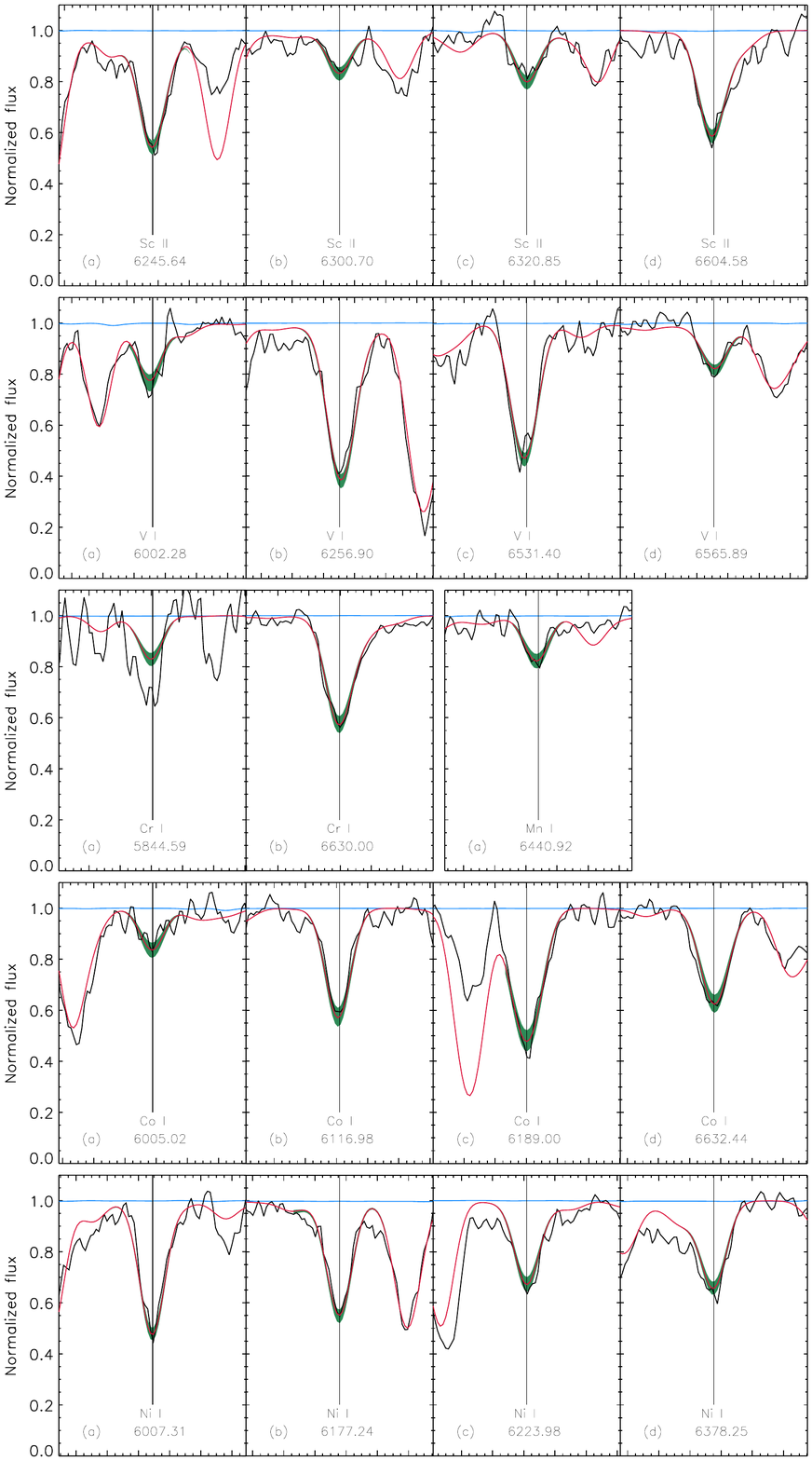}
\caption[B3-f1: spectral line fit]{Stellar lines used for the calculations of the line-by-line abundance scatter as well as the abundance determination in the analysis in the spectrum of B3-f1. Same colour notations and axes as in Figure \ref{arc_spec}. The \ion{Cr}{i} line in panel (a) is strongly affected by the noise and was not used in the final abundance measurement.} \label{b3f_spec}
\end{figure*}

\end{appendix}

\end{document}